\documentclass[twocolumn,english,aps]{revtex4-1}
\usepackage[T1]{fontenc}
\usepackage[latin9]{inputenc}
\usepackage{babel}
\usepackage{units}
\usepackage{amsmath}
\usepackage{graphicx}
\usepackage{esint}
\usepackage[unicode=true,pdfusetitle,
 bookmarks=true,bookmarksnumbered=false,bookmarksopen=false,
 breaklinks=false,pdfborder={0 0 1},backref=false,colorlinks=false]
 {hyperref}
\usepackage{breakurl}

\makeatletter

\providecommand{\tabularnewline}{\\}

\usepackage{babel}

\makeatother

\begin{document}

\title{Transport and localization in a topological phononic lattice with
correlated disorder}

\author{Zhun-Yong Ong}

\email{ongzy@ihpc.a-star.edu.sg}

\affiliation{Institute of High Performance Computing, Singapore}

\author{Ching Hua Lee}

\email{calvin-lee@ihpc.a-star.edu.sg}

\affiliation{Institute of High Performance Computing, Singapore}
\begin{abstract}
Recently proposed classical analogs of topological insulators in phononic
lattices have the advantage of much more accessible experimental realization
as compared to conventional materials. Drawn to their potential practical
structural applications, we investigate how disorder, which is generically
non-negligible in macroscopic realization, can attenuate the topologically
protected edge (TPE) modes that constitute robust transmitting channels
at zero disorder. We simulate the transmission of phonon modes in
a quasi-one-dimensional classical lattice waveguide with mass disorder,
and show that the TPE mode transmission remains highly robust ($\Xi\sim1$)
in the presence of uncorrelated disorder, but diminishes when disorder
is spatially correlated. This reduction in transmittance is attributed
to the Anderson localization of states within the mass disorder domains.
By contrast, non-TPE channels exhibit qualitatively different behavior,
with spatial correlation in the mass disorder leading to significant
transmittance reduction (enhancement) at low (high) frequencies. Our
results demonstrate how TPE modes drastically modify the effect of
spatial correlation on mode localization. 
\end{abstract}
\maketitle

\section{Introduction}

Among the more striking recent advances in acoustic metamaterials
has been the development of a class of engineered metamaterials known
as topological phononic crystals \cite{PWang:PRL15_Topological,ABKhanikaev:NatComm15_Topological,ZYang:PRL15_Topological,mousavi2015topologically,nash2015topological,susstrunk2015observation}.
Like their electronic analogs commonly known as topological insulators
\cite{fu2007,fu2007topological,qi2008,zhang2009,qi2011,hasan2010,lee2015},
they support edge-localized excitations that propagate without significant
attenuation due to their supposed immunity to backscattering by defects.
This peculiar property is a hallmark of topological protection from
nontrivial bulk topological properties in momentum space and potentially
allows the topologically protected edge (TPE) modes to be exploited
for novel applications in phononic circuits and waveguides, where
the high transmission fidelity and the simple linear dependence of
system response on the transfer route are highly beneficial for device
performance \cite{mousavi2015topologically}. The realization of such
systems can lead to improved functionalities for ultrasonic imaging,
sonars, and noise absorbing or enhancing devices. 

However, although it has been demonstrated numerically \cite{PWang:PRL15_Topological,ZYang:PRL15_Topological}
and experimentally \cite{nash2015topological,susstrunk2015observation}
that individual TPE modes can circumvent point or isolated defects,
the propagation of topological modes across a random  medium with
spatially distributed disorder, where the system effectively consists
of heterogeneous domains of possibly distinct topological character,
remains poorly understood, despite their relevance to real phononic
lattices in which structural imperfections may appear. This scenario
is especially relevant to real systems where the phonon wavelength
can be smaller than the disorder domain size. Theoretical studies
of electronic topological insulators in condensed matter physics show
that sufficiently strong disorder can break down the momentum-space
picture that topological protection is built on \cite{li2009topological,groth2009theory,jiang2009numerical,guo2010topological,zhang2012localization,xing2011topological,song2012dependence,xu2012phase,yamakage2011disorder,zhang2013algebraic,girschik2013topological}.
It has been shown numerically by Onoda, Avishai and Nagaosa \cite{MOnoda:PRL07_Localization},
as well as by Castro and co-workers \cite{ECastro:PRB15_Anderson,ECastro:PRB16_Absence},
that localized states start to form in the bulk band gap at high disorder
levels like in conventional Anderson localization \cite{PWAnderson:PR58_Localization}
in two-dimensional systems. Chu, Lu and Shen \cite{RChu:EPL12_Quantum}
also showed that in a quantum spin hall system with a high enough
density of antidots, the breakdown of quantized \emph{electrical}
conduction through the TPE modes is accompanied by the formation of
localized bound states in the bulk band gap. Connected to the disorder-induced
breakdown of TPE states are the numerical results of Li and co-workers
\cite{li2009topological} showing that the electrical conductance
can be quantized \emph{within the conduction band} (CB), instead of
the bulk band gap, when the disorder strength is sufficiently large
to localize the CB bulk modes and create extended edge modes, and
this phenomenon has been called the topological Anderson insulator
(TAI). However, it was discovered that the quantized conductance plateau
in TAIs can be destroyed through coupling between opposite edge modes
by partially delocalized bulk modes \cite{girschik2013topological,girschik2015percolating}
when the disorder is spatially correlated. This suggests that the
transport robustness of TPE modes is sensitive to the \emph{spatial
distribution} as well as the strength of the disorder.

Given the importance of TPE modes in topological phononic crystals
and other metamaterials to the realization to novel device applications,
it is imperative to have a deeper understanding of how they propagate
through disordered media and of the possible suppression of energy
diffusion by Anderson localization \cite{ZShi:OptExp15_Statistics},
a wave phenomenon in both classical and quantum systems. In addition
to its possible relevance to novel acoustic applications, the use
of a classical lattice \cite{HHu:NatPhys08_Localization} to investigate
the phenomenon of Anderson localization in topologically nontrivial
systems also allows us to make direct comparison with experiments
\cite{nash2015topological} where the real space propagation of the
modes can be observed. At the more fundamental level and going beyond
conventional condensed matter physics, the interplay between Anderson
localization and topology has not been fully explored, unlike localization
in simple harmonic lattices which has been studied extensively \cite{MWilliams:PRB85_Numerical,BLi:PRL01_CanDisorder,FABFDeMoura:PRB03_DelocalizationCorrelatedRandomMasses,AChaudhuri:PRB10_PhonLocalization,CMonthus:PRB10_ParticipRatios,SPinski12:JPCM_Localization}.

In this work, we explore the correlated disorder-induced changes in
phonon transmission and the onset of bulk localization in a \emph{multichannel}
phononic Chern insulator lattice waveguide. We apply our recent extension
of the atomistic Green's function method, originally formulated for
studying nanoscale phonon transmission \cite{ZYOng:PRB15_Efficient},
to characterize the transport of topological and non-topological modes
in a disordered environment by analyzing the dependence of the \emph{individual}
phonon mode transmission on frequency, momentum and topology. We show
that correlated disorder can result in the breakdown of robust TPE
mode transmission and this breakdown occurs simultaneously with the
formation of Anderson-localized states within the disordered region.
The spatial distribution of the localized states in different domains
also depends on the relative position of the mode frequency with respect
to the topological band gap edges.

The organization of the paper is as follows. We give a brief overview
of our model two-dimensional Chern insulator phononic lattice and
the emergence of the TPE modes when lattice has a finite width. We
then describe the configuration of the simulated topological phononic
lattice waveguide, which we use to characterize the transmission of
individual modes through a finite disordered region. The detailed
description of the numerical implementation is given in the appendices.
We then present the numerical results and discuss how the effect of
disorder on the bulk and TPE modes. The transmission reduction of
the TPE modes is connected to the change in density of states within
the topological band gap. By comparing the local density of states
and the mass disorder distribution, we show how the spatial distribution
of the localized states depends on the frequency and the type of mass
disorder.

\section{Description of topological phononic crystal and waveguide}

\subsection{\textit{\emph{Topological and bulk modes in phononic Chern insulators}}}

As introduced theoretically in Ref. \cite{PWang:PRL15_Topological}
and demonstrated experimentally in Ref. \cite{nash2015topological},
TPE modes in a two-dimensional (2D) lattice of masses connected by
linear springs can be realized by introducing time-reversal symmetry
breaking via gyroscopic coupling \cite{PWang:PRL15_Topological}.
The phonon modes are described by the eigenvalue equation 
\begin{equation}
[\mathbf{K}(\mathbf{q})-\omega^{2}\mathbf{M}]\mathbf{U}=\mathbf{0},\label{Eq:PhononEigenEqn}
\end{equation}
with $\omega$ and $\mathbf{q}$ being the eigenfrequency and wave
vector. The stiffness ($\mathbf{K}$) and mass ($\mathbf{M}$) matrices
depend on the lattice configuration, while $\mathbf{U}$ represents
an eigenmode. In Eq. (\ref{Eq:PhononEigenEqn}), the phonon lattice
is mathematically described by a tight-binding ``Hamiltonian'' $\mathbf{M^{-1/2}KM^{-1/2}}$,
with each band $n$ possessing a Berry flux $F_{n}^{xy}=-i\left(\langle\partial_{q_{y}}\mathbf{U_{n}}|\partial_{q_{x}}\mathbf{U_{n}}\rangle-\langle\partial_{q_{x}}\mathbf{U_{n}}|\partial_{q_{y}}\mathbf{U_{n}}\rangle\right)$
yielding a Chern number $C_{n}=\frac{1}{2\pi}\int_{BZ}F_{n}^{xy}\;d^{2}\mathbf{q}$.
These fluxes acquire nonzero values in the presence of time-reversal
breaking, and can integrate to nonzero Chern numbers, \emph{i.e.},
give rise to nontrivial topology, with appropriate gyroscopic coupling.

\begin{figure}
\includegraphics[width=8.5cm]{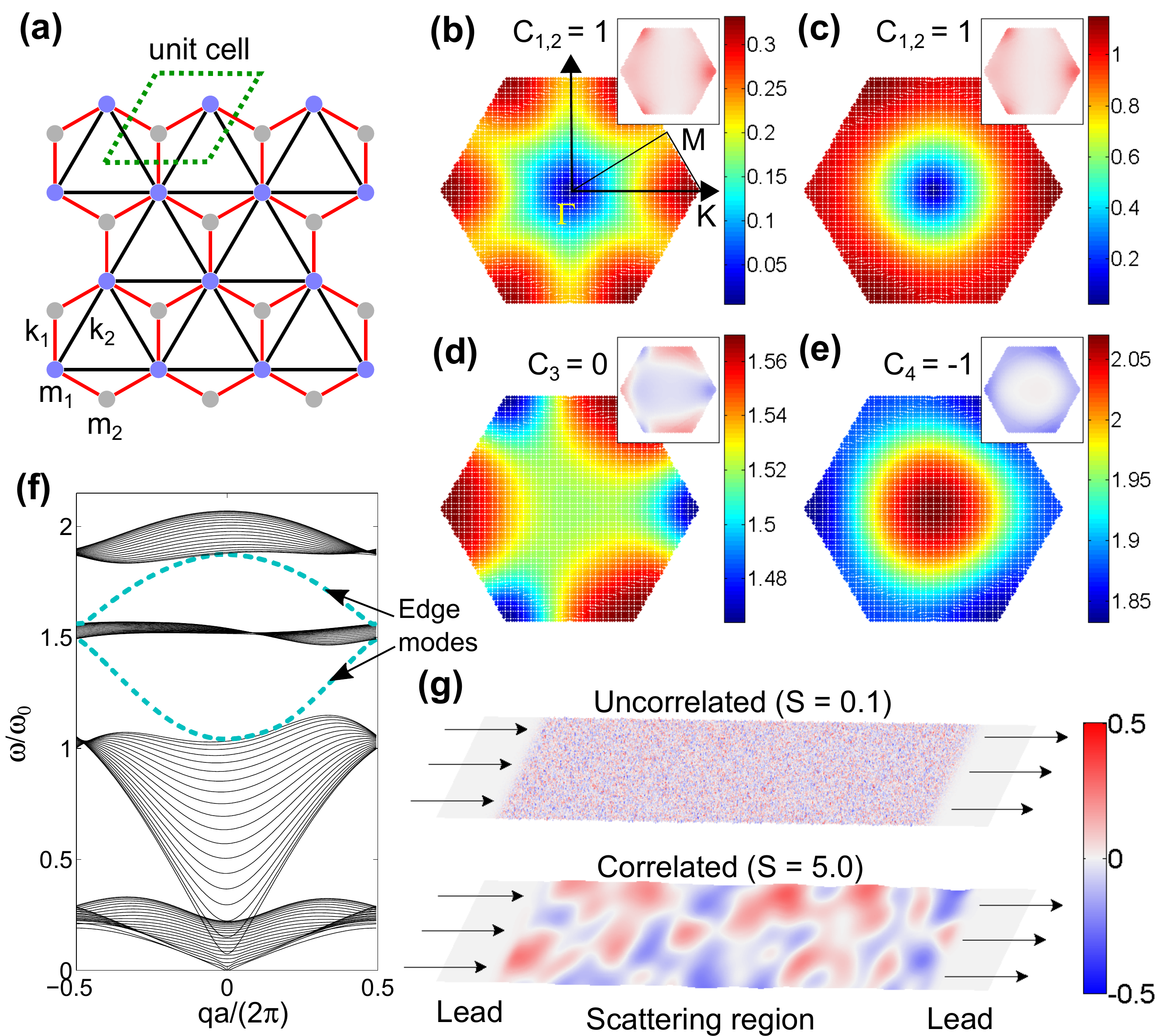}
\caption{(a) Schematic of the hexagonal lattice. The blue (gray) circles represent
the masses $m_{1}$ ($m_{2}$) while the red (black) lines represent
the linear springs $k_{1}$ ($k_{2}$). (b)-(e) Plot of the first
to fourth phonon band over the first Brillouin zone (BZ). The units
of the color scale are in $\omega_{0}$. The accompanying insets show
the Berry fluxes $F_{n}^{xy}$ with their corresponding Chern numbers
($C_{n}$, $n=$ 1 to 4). We combine the Chern number ($C_{1,2}$)
for the first two bands because they are degenerate at the BZ center.
(f) Phonon dispersion for $N_{W}=20$ lattice waveguide. (g) Schematic
of uncorrelated and correlated mass disorder in a waveguide. The color
scale for $\delta m(\mathbf{r})$ is shown. }

\label{Fig:PhononBands} 
\end{figure}

A prototypical 2D topological phononic lattice is given by the honeycomb
model from Ref. \cite{PWang:PRL15_Topological}, which we use in our
simulations {[}Fig \ref{Fig:PhononBands}(a){]}. It consists of identical
masses $m$ with only in-plane motion connected by nearest and next-nearest
neighbor springs {[}Fig. \ref{Fig:PhononBands}(a){]}. Since there
are two masses per unit cell and two polarizations ($x$ and $y$),
Eq. (\ref{Eq:PhononEigenEqn}) yields two acoustic and two optical
phonon bands {[}Fig. \ref{Fig:PhononBands}(b) to (e){]}. We define
the characteristic frequency $\omega_{0}=\sqrt{k_{1}/\langle m\rangle}$
where $k_{1}$ and $\langle m\rangle$ are the nearest-neighbor spring
constant and the average mass, respectively. Robust TPE modes are
formed in the gap between these topological bands {[}Fig. \ref{Fig:PhononBands}(f){]}
when the 2D lattice is terminated by edges like in a waveguide {[}Fig.
\ref{Fig:PhononBands}(g){]} where TPE modes of opposite momentum
are localized at either one of the edges separated by $N_{W}$ unit
cells in the transverse direction. Figure \ref{Fig:PhononBands}(f)
shows the phonon dispersion for a pristine $N_{W}=20$ armchair-edge
lattice waveguide, identical to the one in Ref. \cite{PWang:PRL15_Topological}.

\subsection{Topological phononic lattice waveguide configuration}

To simulate the transport of TPE and non-TPE modes, we set up the
lattice waveguide with a finite mass-disordered scattering region
of length $N_{L}a$ sandwiched between two pristine semi-infinite
leads. $N_{L}$ is the number of unit cells spanning the scattering
region and $a$ is the 1D lattice constant. Disorder in the scattering
region is introduced by modifying the mass $m(\mathbf{r})$ at each
site $\mathbf{r}$ by a random variable $\delta m(\mathbf{r})$, \emph{i.e.},
$m(\mathbf{r})=\langle m\rangle+\delta m(\mathbf{r})$ where $\langle m\rangle=1.0$.
The spatial correlation in $\delta m(\mathbf{r})$ is characterized
by the function 
\begin{equation}
\langle\delta m(\mathbf{r}_{1})\delta m(\mathbf{r}_{2})\rangle=\langle\delta m^{2}\rangle\exp[-|\mathbf{r}_{1}-\mathbf{r}_{2}|^{2}/(2S^{2})]\ ,\label{Eq:MassFluctCorrel}
\end{equation}
where $S$ is the correlation length and $\langle\ldots\rangle$ is
the ensemble average. The detailed procedure for generating the correlated
disorder in Eq. (\ref{Eq:MassFluctCorrel}) is described in Appendix
\ref{Append:GenerateCorrelatedMass}. We note that the mass disorder
results in the changes in the diagonal and off-diagonal matrix elements
of the ``Hamiltonian'' $\mathbf{M^{-1/2}KM^{-1/2}}$. Such mass disorder
in the scattering region can be realized experimentally by using `atoms'
of different masses in the lattice \cite{PWang:PRL15_Topological,nash2015topological}.
The mass $m(\mathbf{r})$ is treated as a continuous random variable
like in other studies of disordered harmonic lattices \cite{BLi:PRL01_CanDisorder,AChaudhuri:PRB10_PhonLocalization,SPinski12:JPCM_Localization}.
We set the nearest-neighbor distance between the masses $r_{\textrm{NN}}$
to $1.0$. We take $S=0.1$ to represent uncorrelated disorder, since
$S\ll r_{\textrm{NN}}$, and $S=5.0$ to represent correlated disorder.
The root-mean-square mass disorder is set as $\sqrt{\langle\delta m^{2}\rangle}=0.1$.
Figure \ref{Fig:PhononBands}(g) shows the spatial profile of $\delta m(\mathbf{r})$
for $S=0.1$ and $S=5.0$, with the latter showing significant domains
of positive ($\delta m>0$, in red) or negative ($\delta m<0$, in
blue) mass disorder. In the waveguide, incoming left lead phonons
are either reflected from or scattered across the disordered scattering
region to the available right-lead channels. Due to transverse subband
quantization, there are  $N(\omega)$ transmitting and receiving channels
at each frequency. The calculated \emph{transmission coefficient}
(TC) of each left-lead mode, $\Xi_{n}(\omega)$ for $n=1,\ldots,N(\omega)$,
gives the fraction of energy that is transmitted after scattering.
We define the transmittance $T(\omega)$ as the sum of the TCs at
each frequency, \emph{i.e.} $T(\omega)=\sum_{n=1}^{N(\omega)}\Xi_{n}(\omega)$.

\section{Results and discussion}

\subsection{\textit{\emph{Effect of disorder correlation on transmission coefficients
}}}

\begin{figure}
\includegraphics[width=8cm]{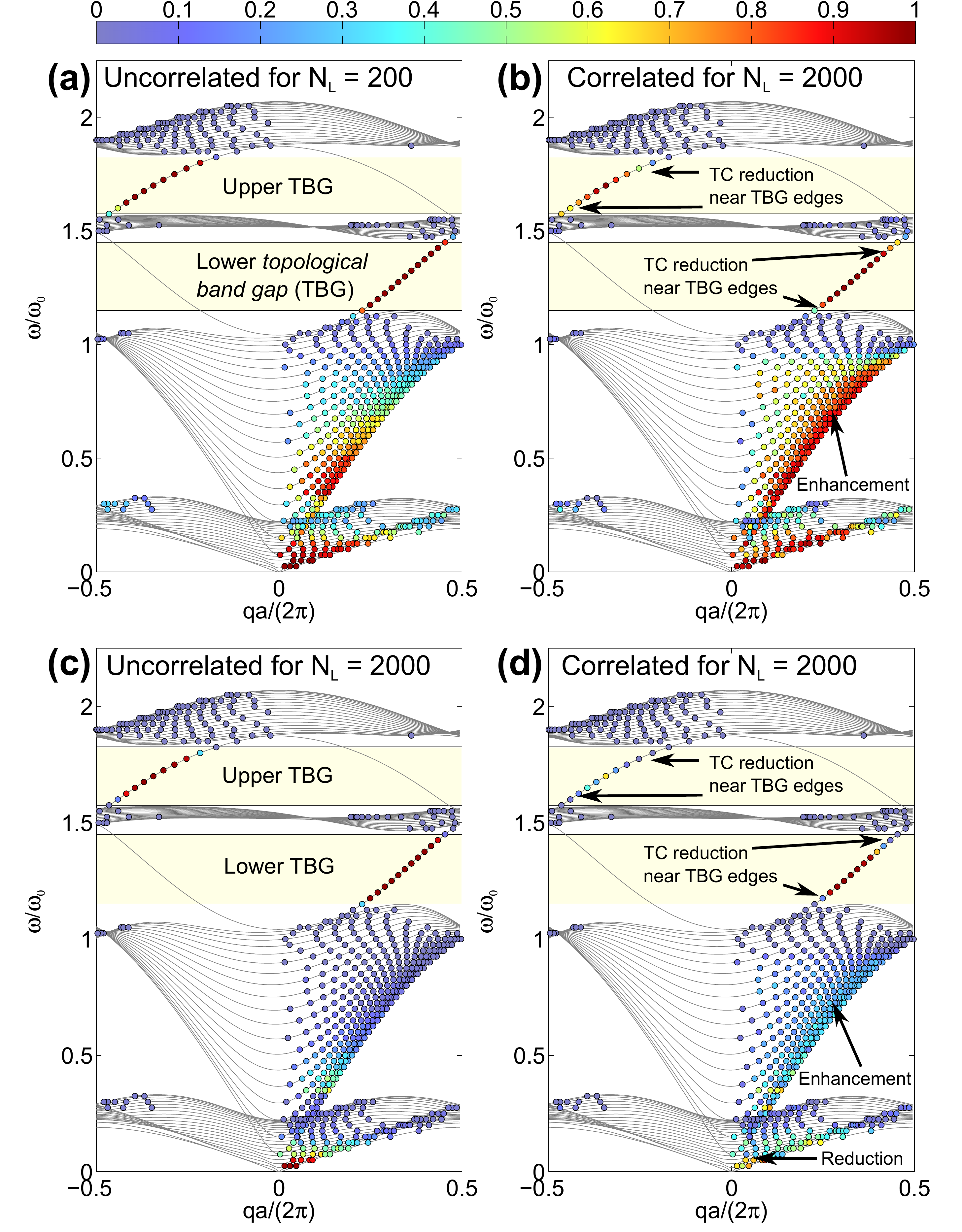}

\begin{tabular}{|c|c|c|c|}
\hline 
Phonon modes  & \multicolumn{3}{c|}{Phonon mode transmission $\Xi_{n}$}\tabularnewline
\hline 
 & Higher $q$  & Higher $\omega$  & Higher $S$\tabularnewline
\hline 
\hline 
Low-frequency  & Enhanced  & Reduced  & Reduced \tabularnewline
(bulk)  &  &  & \tabularnewline
\cline{1-1} \cline{4-4} 
High-frequency  &  &  & Enhanced \tabularnewline
(bulk)  &  &  & \tabularnewline
\hline 
Topologically  & \multicolumn{2}{l|}{Enhanced, if mode shifts } & Reduced \tabularnewline
protected edge  & \multicolumn{2}{l|}{away from TBG edge} & \tabularnewline
\hline 
\end{tabular}

\caption{Average phonon mode transmission coefficients for $N_{L}=200$ with
(a) uncorrelated and (b) correlated disorder. The transmission coefficients
have a value between 0 and 1, and are indicated by color according
to the top color bar scale. The effects of correlated disorder on
the transmission attenuation can be seen more clearly for $N_{L}=2000$
with (c) uncorrelated and (d) correlated disorder. The yellow-shaded
bands represent the topological band gaps. Since the propagation direction
is rightward, only modes with a positive group velocity ($\partial\omega/\partial q>0$)
can be transmitted. The table summarizes the effects of changing the
wave vector ($q$), frequency ($\omega$) and disorder correlation
length ($S$) on the transmission coefficients $\Xi_{n}$ for different
phonon modes. More plots for $S=1.0$ and $2.0$ are available in
Appendix \ref{Append:AdditionalTransmissionData}. }

\label{Fig:PhononDispersionAndTransmission} 
\end{figure}

We first consider the case of uncorrelated disorder ($S=0.1$). Since
translational symmetry is still preserved after disorder averaging,
the Chern number remains well-defined in momentum space and the TPE
states are robust. This is evident from Figs. \ref{Fig:PhononDispersionAndTransmission}
(a) and (c), which show the phonon dispersion with the computed average
mode transmission coefficients superimposed on it. The TC is close
to unity for the TPE modes in the two yellow-shaded frequency ranges,
each of which delineates a ``topological band gap'' (TBG). The lower
TBG ($1.15<\omega/\omega_{0}<1.45$) lies between the second and third
bands {[}Figs. \ref{Fig:PhononBands}(c) and (d){]} and the upper
TBG ($1.575<\omega/\omega_{0}<1.825$) lies between the third and
fourth bands {[}Figs. \ref{Fig:PhononBands}(d) and (e){]}. The topological
mode TCs show no appreciable decrease as the waveguide length is increased
ten times from $N_{L}=200$ to $2000$, attesting to their immunity
to backscattering.

The TCs for the bulk (non-TPE) modes with uncorrelated disorder exhibit
the following universal behavior. Notably, the TC in each branch decreases
at higher frequencies, a tendency also observed in disordered single-channel
atomic chains \cite{HMatsuda:PTP70_Localization,ADhar:PRL01_DisorderedHarmonicChain,ZYOng:JPCM14_Ballistic,ZYOng:PRB14_Enhancement}.
As expected, the TCs decrease as the size of the system increases
from $N_{L}=200$ {[}Fig. \ref{Fig:PhononDispersionAndTransmission}(a){]}
to $2000$ {[}Fig. \ref{Fig:PhononDispersionAndTransmission}(c){]},
indicating that transmission is attenuated by the length of the intervening
disordered medium. More interestingly, at the same frequency, the
TC is generally lower for modes nearer to the BZ center ($q=0$) and
with smaller group velocities, indicating that slower modes are more
strongly attenuated by disorder. 

When spatial correlation {[}Eq. (\ref{Eq:MassFluctCorrel}){]} is
introduced in the disorder, the system effectively becomes a conglomerate
of ``islands'' with different masses, and translational symmetry
is broken even after disorder averaging. Consequently, the TCs for
non-TPE modes are enhanced (reduced) at higher (lower) frequencies
below the bottom edge of the lower TBG, \emph{i.e.}, for $\omega/\omega_{0}<1.15$.
This is evident from comparing Figs. \ref{Fig:PhononDispersionAndTransmission}(b)
and (d), where $S=5.0$, with Figs. \ref{Fig:PhononDispersionAndTransmission}(a)
and (c) with $S=0.1$. Additional results showing the change in TC
with correlation length are given in Appendix \ref{Append:AdditionalTransmissionData}.
The most striking effect of disorder correlation is the transmission
attenuation of states that are topologically protected when the disorder
is uncorrelated. In Figs. \ref{Fig:PhononDispersionAndTransmission}(c)
and (d), we observe considerable reduction of the TC in former TPE
states within the upper TBG, with the reduction more pronounced near
the band gap edges. This is related to the breakdown of a single bulk
Chern number and the onset of bulk localization. 

\begin{figure}
\includegraphics[width=8cm]{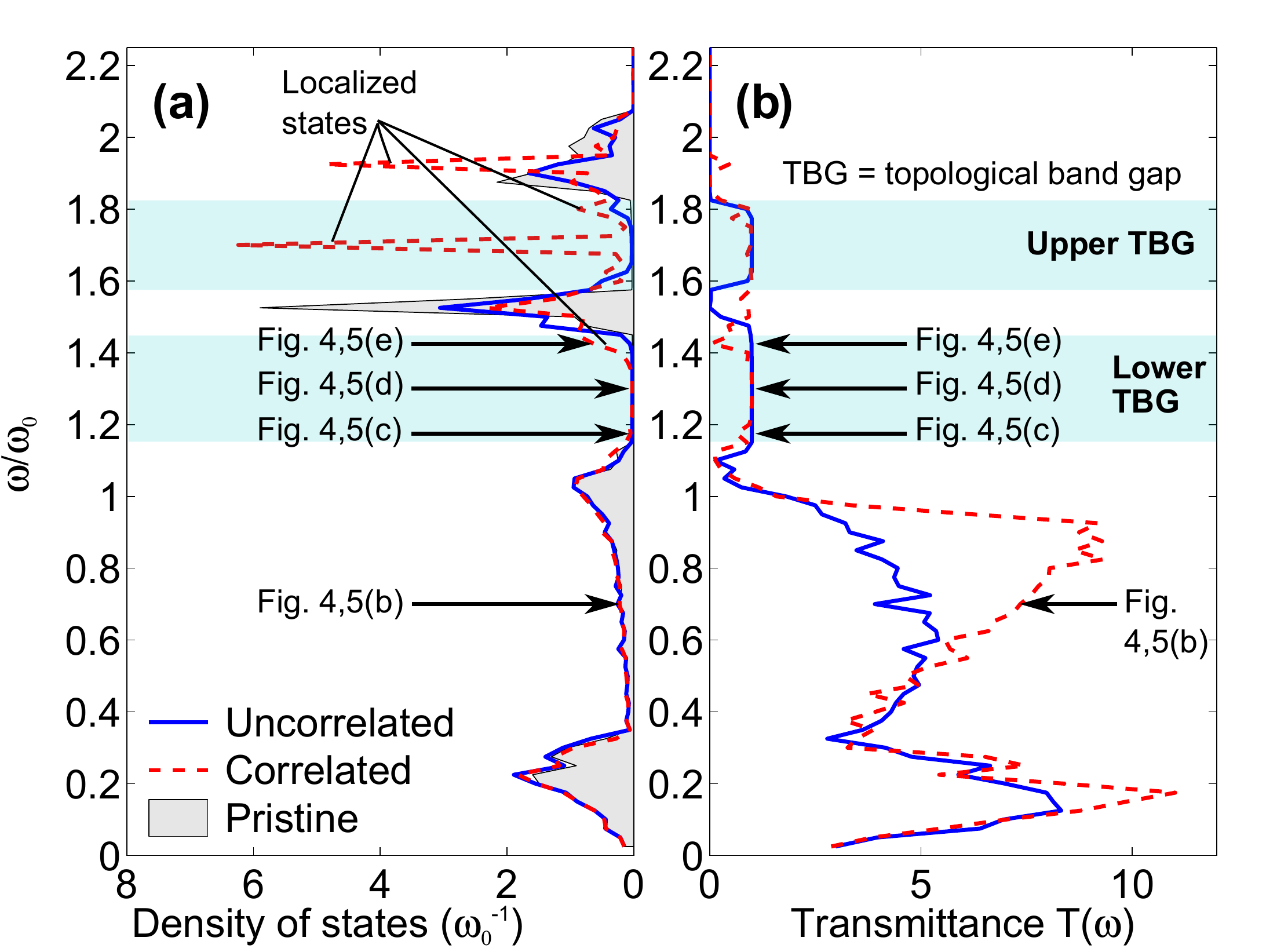} \caption{(a) Total density of states $\overline{\rho}(\omega)$ spectra for
the scattering region in the $N_{L}=200$ waveguide with uncorrelated
(solid line), correlated (dashed line) and no disorder (pristine).
(b) The corresponding transmittance spectra for uncorrelated and correlated
disorder. The corresponding local density of states for $\omega/\omega_{0}$
= 0.700, 1.175, 1.300 and 1.425 are shown in Figs. \ref{Fig:PhononLDOS_uncorrelated}
and \ref{Fig:PhononLDOS_correlated}(b)-(e) and labeled accordingly. }

\label{Fig:PhononDOSAndTransmission} 
\end{figure}

\subsection{\textit{\emph{Changes in transmission coefficients and density of
states}}}

\begin{figure}
\includegraphics[width=8cm]{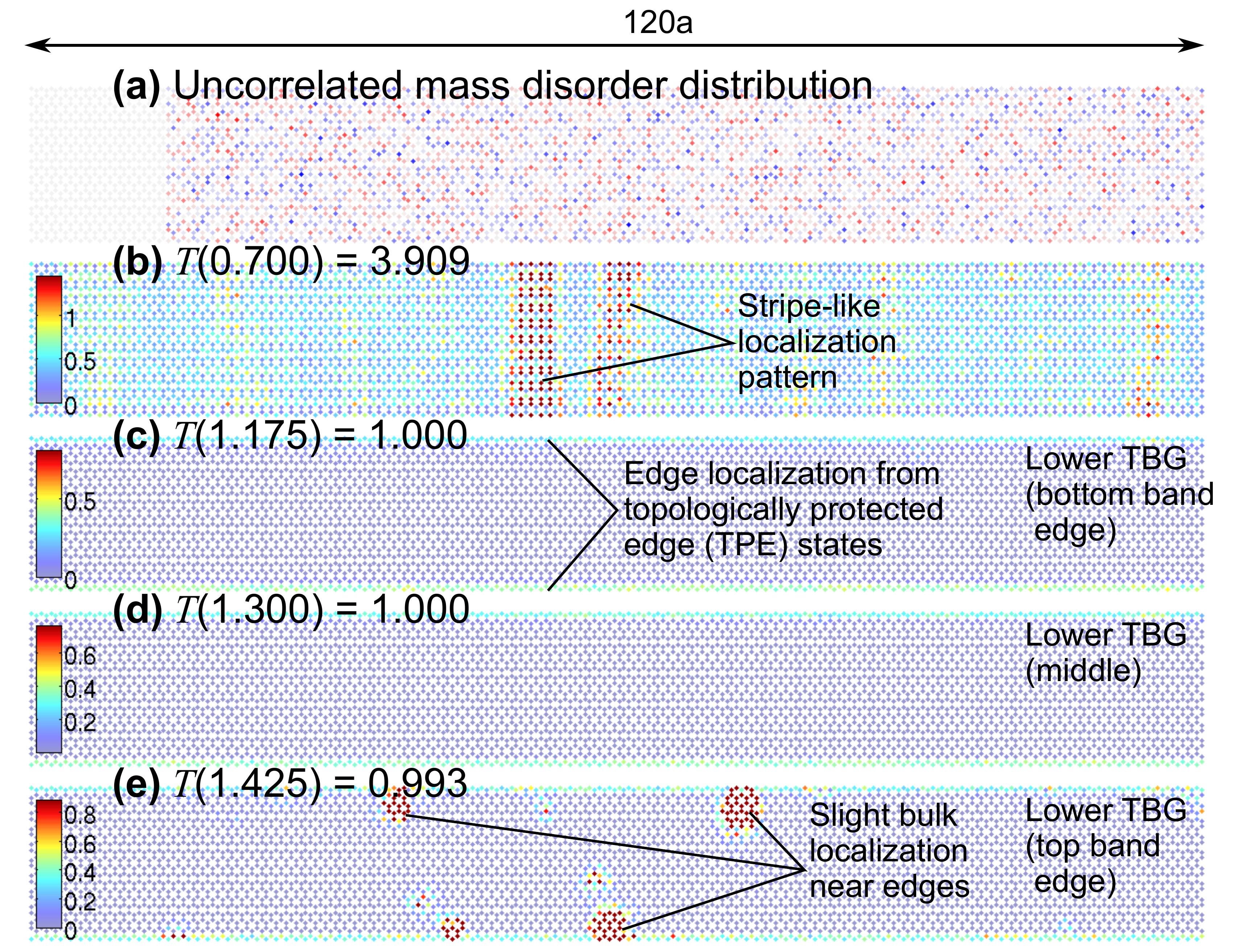} \caption{(a) Uncorrelated mass disorder distribution in a portion of an $N_{L}=200$
waveguide with positive (red) and negative (blue) $\delta m$. Part
of the left lead is also shown. The shading intensity corresponds
to the amplitude of the $\delta m(\mathbf{r})$. (b)-(e) The local
density of states for $\omega/\omega_{0}=$ 0.700, 1.175, 1.300 and
1.425 are shown together with their transmittance $T(\omega)$ values.
(c)-(e) correspond to modes in the lower TBG. Bulk localization only
occurs in the bulk window {[}(b){]}, and to a small extent near the
TBG {[}(e){]}. }

\label{Fig:PhononLDOS_uncorrelated} 
\end{figure}

\begin{figure}
\includegraphics[width=8cm]{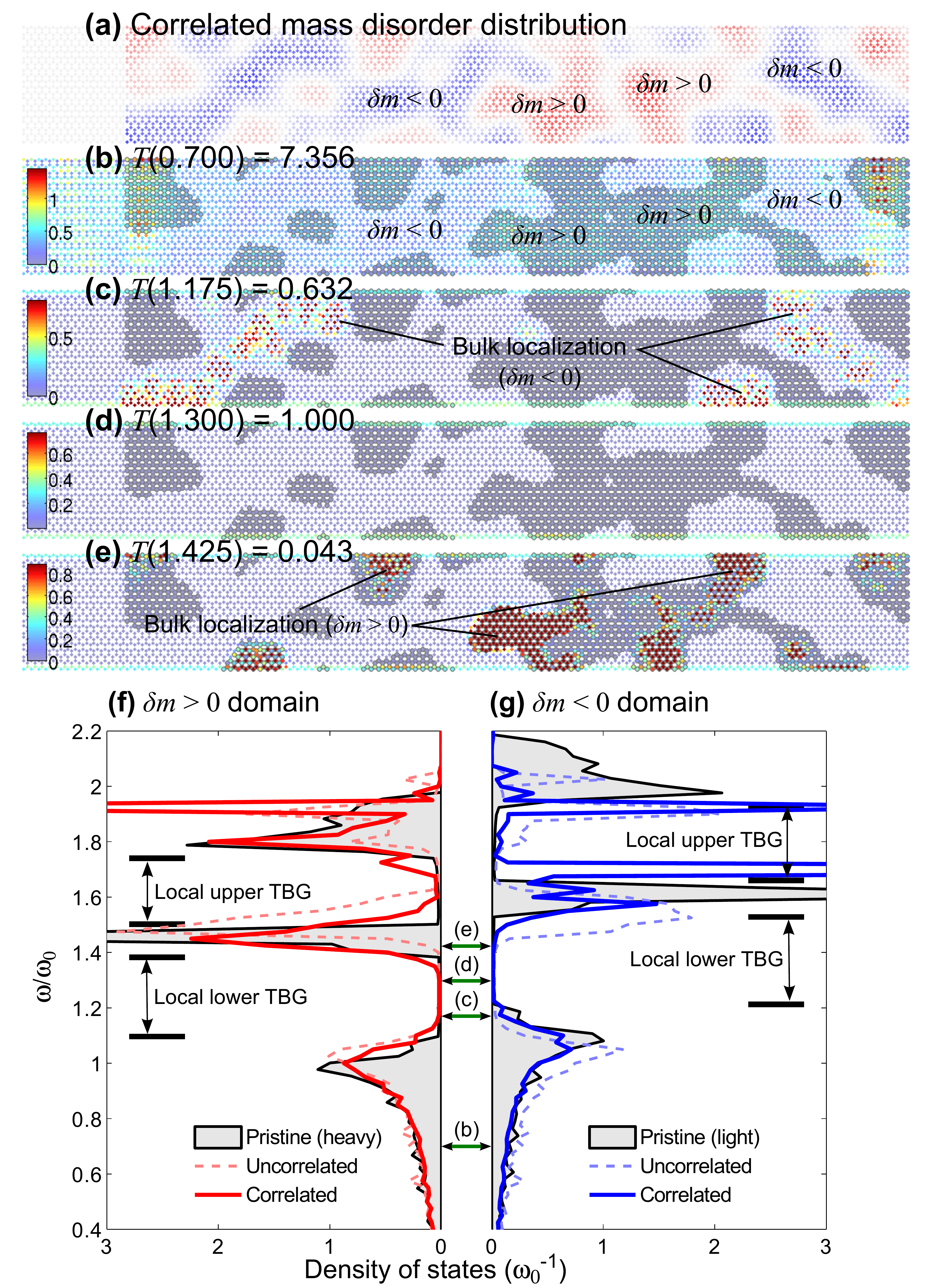} \caption{(a) Plot of $\delta m(\mathbf{r})$ like in Fig. \ref{Fig:PhononLDOS_uncorrelated}
but for a realization of correlated disorder, with positive (red)
and negative (blue) $\delta m$. (b)-(e) The local density of states
(LDOS) at $\omega/\omega_{0}=$ 0.700, 1.175, 1.300 and 1.425 are
shown together with their transmittance $T(\omega)$ values. The sites
in (b)-(e) corresponding to $\delta m>0$ are outlined with a solid
gray line in the background. In the bulk window, the LDOS is well-localized
within the bulk like in Fig. \ref{Fig:PhononLDOS_uncorrelated}(b),
but there is no direct correspondence between the localized LDOS regions
and the mass disorder domains. (f) The LDOS averaged over sites with
$\delta m>0$ for uncorrelated (dashed line) and correlated (solid
line) disorder, and the TDOS for the pristine (heavy) system with
a mass shift of $\langle m\rangle\rightarrow\langle m\rangle+\sqrt{\langle\delta m^{2}\rangle}$.
(g) The LDOS averaged over sites with $\delta m<0$ for uncorrelated
and correlated disorder, and the TDOS for the pristine (light) system
with a mass shift of $\langle m\rangle\rightarrow\langle m\rangle-\sqrt{\langle\delta m^{2}\rangle}$.
The corresponding shifted \emph{local} TBGs for (f) and (g) are also
indicated.}

\label{Fig:PhononLDOS_correlated} 
\end{figure}

The topological mode TC reduction from correlated disorder suggests
coupling between the propagating states from the leads and the localized
states within the scattering region. It is known that the amplitude
of a mode localized within the the interior of a disordered region
is exponentially small at the boundary and couples weakly to the surrounding
degrees of freedom at the boundaries \cite{ZShi:OptExp15_Statistics},
reducing the probability of an incoming propagating mode being transmitted
across the disordered scattering region. To clarify the relationship
between correlated disorder and the formation of the localized states,
we compare the local density of states (LDOS) $\rho(\omega,\mathbf{r})$,
defined in \ref{SubAppend:LDOS}, for both uncorrelated and correlated
disorder (Figs. \ref{Fig:PhononLDOS_uncorrelated} and \ref{Fig:PhononLDOS_correlated}
respectively) at the specific frequencies indicated in Fig. \ref{Fig:PhononDOSAndTransmission},
in which their total density of states (TDOS) $\overline{\rho}(\omega)$,
defined as 
\begin{equation}
\overline{\rho}(\omega)=\frac{\int d\mathbf{r^{\prime}}\rho(\omega,\mathbf{r^{\prime}})\sum_{\mathbf{r}}\delta(\mathbf{r^{\prime}}-\mathbf{r})}{d\omega\int d\mathbf{r^{\prime}}\rho(\omega,\mathbf{r^{\prime}})\sum_{\mathbf{r}}\delta(\mathbf{r^{\prime}}-\mathbf{r})}\label{Eq:TDOS_expression}
\end{equation}
where $\rho(\omega,\mathbf{r})$ is given in Eq. (\ref{Eq:LocalDOS})
and the summation $\sum_{\mathbf{r}}\ldots$ is over lattice degrees
of freedom within the scattering region, and corresponding transmittance
spectra $T(\omega)$ are shown for a single $N_{L}=200$ realization.
The TDOS $\overline{\rho}(\omega)$ corresponds to the normalized
average LDOS within the disordered region and satisfies the relation
$\int d\omega\overline{\rho}(\omega)=1$. 

As expected, uncorrelated disorder localizes bulk states, leading
to reduced bulk transmission, but leaves the TPE states unattenuated.
At $\omega/\omega_{0}=0.700$ in the bulk transmission window, the
LDOS is well-localized with stripe-like patterns within the bulk of
the waveguide because of the uncorrelated disorder {[}Fig. \ref{Fig:PhononLDOS_uncorrelated}(b){]}.
However, near the bottom {[}Fig. \ref{Fig:PhononLDOS_uncorrelated}(c){]}
and middle {[}Fig. \ref{Fig:PhononLDOS_uncorrelated}(d){]} of the
lower TBG, only TPE modes exist and the LDOS is consistently edge-localized
in spite of the disorder. There is some accumulation of bulk-localized
states nearer to the top edge of the lower TBG {[}Fig. \ref{Fig:PhononLDOS_uncorrelated}(e){]},
but that is not sufficient to destroy the near-perfect transmission.
We associate the absence of significant bulk localization in the TBGs
{[}Figs. \ref{Fig:PhononLDOS_uncorrelated}(c)-(e){]} with near-perfect
transmittance ($T(\omega)\ge0.99$). 

When disorder is spatially correlated {[}Fig. \ref{Fig:PhononLDOS_correlated}(a){]},
there exists heterogeneous regions of mass domains that are large
enough to be topologically distinct. This is most apparent in the
LDOS spectra within the TBGs, where bulk states do not exist in the
pristine case. But here, near the bottom of the lower TBG {[}Fig.
\ref{Fig:PhononLDOS_correlated}(c){]}, parts of the LDOS are localized
inside the bulk and confined within the $\delta m<0$ domains delineated
by the unshaded portions of Fig. \ref{Fig:PhononLDOS_correlated}(c),
implying that the bulk-localized states are formed within the $\delta m<0$
domains. At $\omega/\omega_{0}=1.300$ in the middle of the lower
TBG {[}Fig. \ref{Fig:PhononLDOS_correlated}(d){]}, there is no bulk
localization although it is again observed near the top of the lower
TBG {[}Fig. \ref{Fig:PhononLDOS_correlated}(e){]}. However, unlike
Fig. \ref{Fig:PhononLDOS_correlated}(c), the contiguous bulk-localized
portions are confined within the $\delta m>0$ domains. The difference
between the LDOS spectra in Figs. \ref{Fig:PhononLDOS_correlated}(c)-(e)
suggests that the correspondence in the spatial distribution of the
bulk-localized states and the mass disorder domains is frequency-sensitive.

\subsection{\textit{\emph{Bulk localization and local topological band gap shifts}}}

Having established the basic picture of how bulk localization {[}Figs.
\ref{Fig:PhononLDOS_correlated}(b)-(e){]} depends on domain distribution
{[}Fig. \ref{Fig:PhononLDOS_correlated}(a){]}, we connect it to the
change in density of states and transmittance in Figs. \ref{Fig:PhononDOSAndTransmission}(a)
and (b). It is heuristically useful to interpret the $\delta m>0$
($\delta m<0$) domains as islands with positive (negative) mass `doping'
that downshifts (upshifts) the local TBGs. At the bottom edge of the
lower TBG {[}Fig. \ref{Fig:PhononLDOS_correlated}(c){]}, the states
in the $\delta m>0$ domains are in the downshifted \emph{local} TBG
and are thus `TPE-like' while the states in the $\delta m<0$ domains
are outside of the upshifted \emph{local} TBG and can be described
as `bulk-like'. Thus, the bulk-localized states in Fig. \ref{Fig:PhononLDOS_correlated}(c)
only occur in the $\delta m<0$ domains. Similarly, at the top edge
of the lower TBG {[}Fig. \ref{Fig:PhononLDOS_correlated}(e){]}, the
states in the $\delta m>0$ domains are just above the top edge of
the downshifted \emph{local} TBG and `bulk-like' while the states
in the $\delta m<0$ domains are inside the upshifted \emph{local}
TBG and `TPE-like'. Hence, the bulk localization {[}Fig. \ref{Fig:PhononLDOS_correlated}(e){]}
occurs only in the $\delta m>0$ domains. 

To illustrate this explanation, we plot in Fig. \ref{Fig:PhononLDOS_correlated}(f)
the LDOS averaged over the heavier $\delta m>0$ sites for correlated
{[}Figs. \ref{Fig:PhononLDOS_uncorrelated}(a){]} and uncorrelated
{[}Fig. \ref{Fig:PhononLDOS_correlated}(a){]} disorder. The spectra
are similar at low frequencies but diverge as $\omega\rightarrow\infty$.
In particular, the correlated-disorder LDOS gap is downshifted with
respect to the TBGs in Fig. \ref{Fig:PhononDOSAndTransmission}(a).
We interpret the shift as the \emph{local} TBG downshift caused by
positive mass loading. To confirm this interpretation, we plot the
TDOS for a heavy pristine system with a positive mass shift of $\langle m\rangle\rightarrow\langle m\rangle+\sqrt{\langle\delta m^{2}\rangle}$.
The TDOS is much better aligned to the LDOS for correlated ($\delta m>0$)
disorder than for uncorrelated disorder, reinforcing the idea that
the gap downshift is due to the mass loading of the local modes within
the $\delta m>0$ domains. Likewise, we also plot the LDOS averaged
over the lighter $\delta m<0$ sites in Figs. \ref{Fig:PhononLDOS_uncorrelated}(a)
and \ref{Fig:PhononLDOS_correlated}(a) together with the TDOS for
a light pristine system with a negative mass shift of $\langle m\rangle\rightarrow\langle m\rangle-\sqrt{\langle\delta m^{2}\rangle}$.
Similarly, the light pristine TDOS is much better aligned to the average
LDOS for correlated ($\delta m<0$) disorder than for uncorrelated
disorder. Finally, in contrast, the gaps in the LDOS spectra for uncorrelated
disorder $\delta m>0$ {[}Fig. \ref{Fig:PhononLDOS_correlated}(f){]}
and $\delta m<0$ {[}Fig. \ref{Fig:PhononLDOS_correlated}(g){]} are
well-aligned to each other and to the TBGs in Fig. \ref{Fig:PhononDOSAndTransmission}(a),
confirming that mass loading has less effect on their TBGs.

\section{Summary and conclusions}

We have studied the transmission of topological protected edge (TPE)
and non-topological bulk modes through a finite mass-disordered lattice
waveguide and found that correlated mass disorder enhances (reduces)
the transmission of high (low) frequency non-topological modes. However,
the transmission of TPE modes near the band edges of topological band
gap (TBG) is degraded by correlated disorder because of the formation
of bulk-localized states in topologically distinct mass ($\delta m>0$
and $\delta m<0$ ) domains in which one can effectively define a
shifted \emph{local} topological band gap. We find that bulk localization
in the mass domain is only permitted if the mode frequency lies outside
of the local TBG. This suggests that we can control the spatial localization
of acoustic energy in a topological phononic crystal through correlated
disorder. 
\begin{acknowledgments}
We acknowledge financial support from the Agency for Science, Technology
and Research (Singapore).
\end{acknowledgments}

\appendix

\section{Generation of correlated mass disorder \label{Append:GenerateCorrelatedMass}}

\begin{figure*}
\includegraphics[width=5cm]{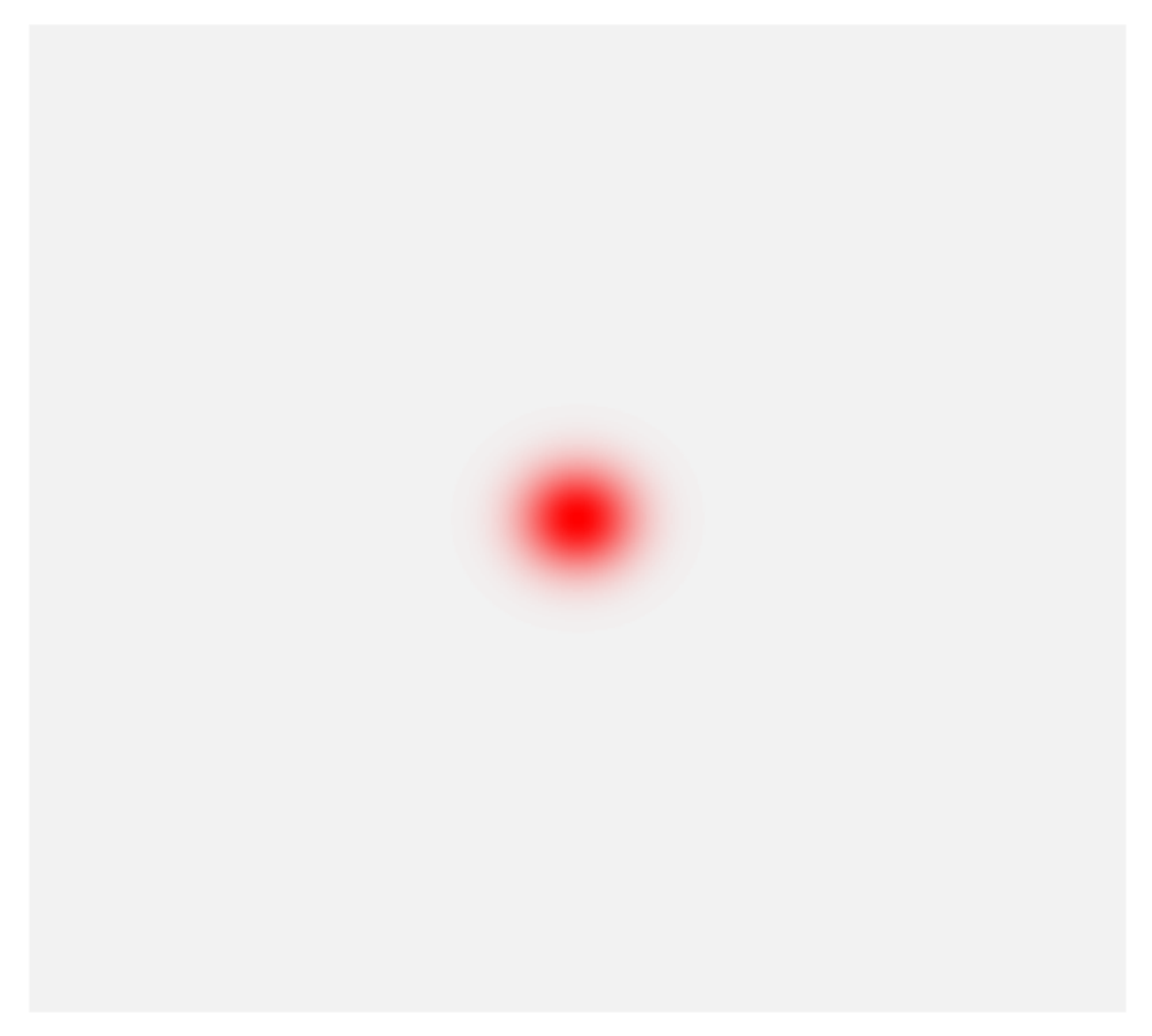}\includegraphics[width=5cm]{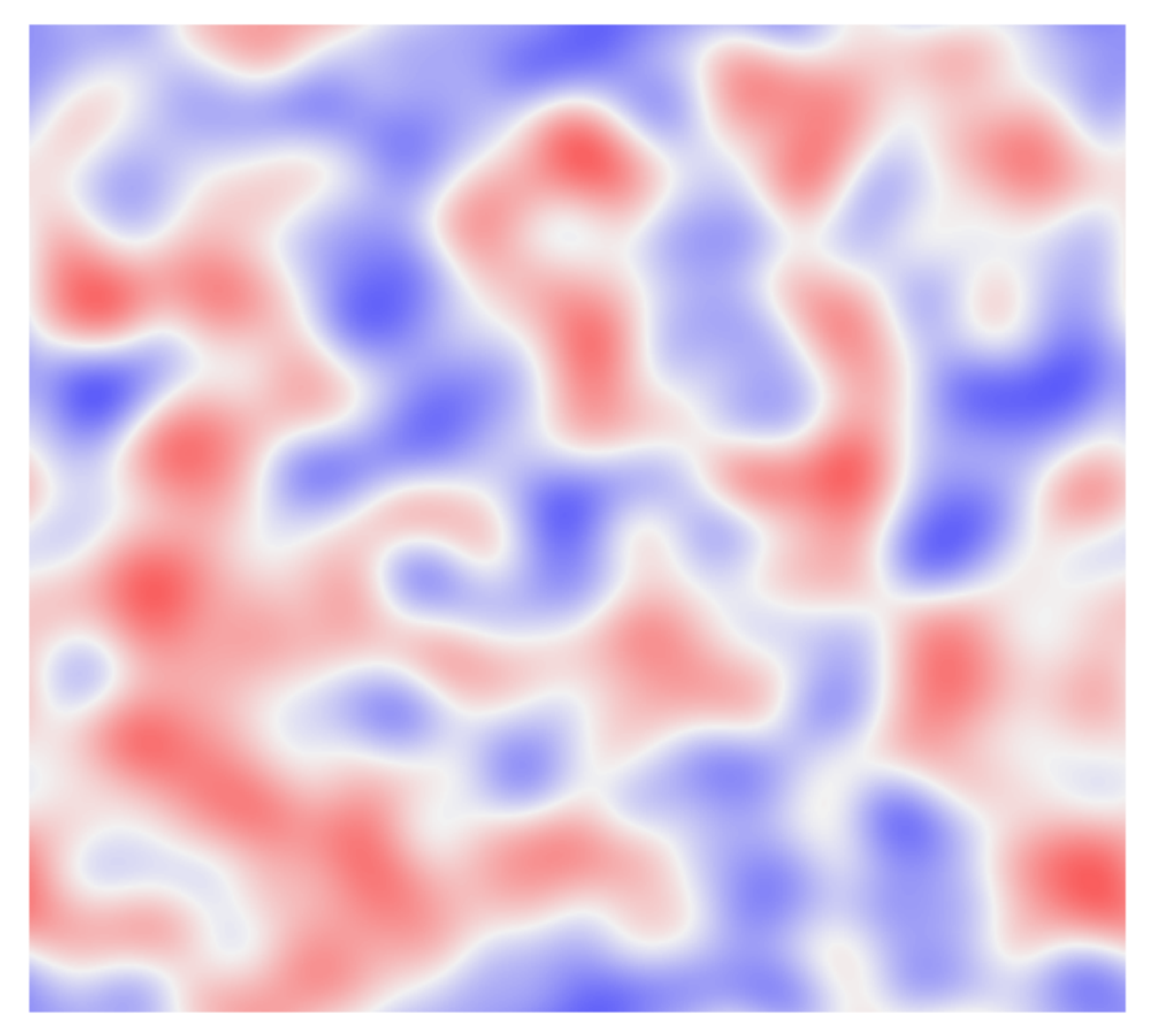}

\caption{The left shows the function $f(x,y)=\frac{1}{\pi S^{2}}\exp\left(-\frac{x^{2}+y^{2}}{S^{2}}\right)$
from Eq.~(\ref{Eq:GaussianFunction}) while the right shows the function
$h(x,y)$ from Eq.~(\ref{Eq:CorrelatedRandomFn}) for $S=5.0$.}
\label{Fig:GaussianAutocorr} 
\end{figure*}

We first generate a dense two-dimensional (2D) Cartesian grid in the
$x$-$y$ plane for the Gaussian function 
\begin{equation}
f(x,y)=\frac{1}{\pi S^{2}}\exp\left(-\frac{x^{2}+y^{2}}{S^{2}}\right)\label{Eq:GaussianFunction}
\end{equation}
over the domain where $-\frac{D}{2}<x\leq\frac{D}{2}$ and $-\frac{D}{2}<y\leq\frac{D}{2}$
for $D\gg S$. The Fourier components $\tilde{f}(k_{x},k_{y})=\mathcal{F}[f(x,y)]$
are computed by taking the discrete Fourier transform. We then multiply
each Fourier component by a random phase factor $\theta(k_{x},k_{y})$
uniformly distributed between $0$ and $2\pi$:$\tilde{f}(k_{x},k_{y})\rightarrow\tilde{f}(k_{x},k_{y})e^{i\theta(k_{x},k_{y})}$.
The 2D random function $h(x,y)$ is obtained by taking the real part
of the inverse Fourier transform of $\tilde{f}(k_{x},k_{y})e^{i\theta(k_{x},k_{y})}$,
\emph{i.e.}, 
\begin{equation}
h(x,y)=\sqrt{2}\text{Re}\mathcal{F}^{-1}[\tilde{f}(k_{x},k_{y})e^{i\theta(k_{x},k_{y})}]\ .\label{Eq:CorrelatedRandomFn}
\end{equation}
Figure~\ref{Fig:GaussianAutocorr} shows the spatial profile of $f(x,y)$
and $h(x,y)$, with the latter displaying distinct domains. It can
be shown that the autocorrelation function of $h(x,y)$ has a Gaussian
form, \emph{i.e.}, 
\[
\frac{\langle h(x,y)h(0,0)\rangle}{\langle h(0,0)^{2}\rangle}=\exp\left(-\frac{x^{2}+y^{2}}{2S^{2}}\right)\ ,
\]
where $\langle\ldots\rangle$ represents the average taken over all
disorder realizations. Therefore, the mass disorder at site $\mathbf{r}=(x,y)$
is given by 
\begin{equation}
\delta m(\mathbf{r})=\sqrt{\frac{\langle\delta m\rangle}{\langle h(0,0)^{2}\rangle}}h(x,y)\ .\label{Eq:RandomMassDisorder}
\end{equation}
Roughly speaking, the length scale of the domains in $h(x,y)$ is
$\sim2S$.

\begin{widetext}

We note that the above approach can be generalized to arbitrarily
correlated textures. Suppose we start from a generic distribution
$f(\mathbf{r})$. We want to derive the spatial correlation of $h(\mathbf{r})=\sqrt{2}\text{Re}\mathcal{F}^{-1}[\tilde{f}(\mathbf{k})e^{i\theta(\mathbf{k})}]$,
i.e. the inverse Fourier transform of the product of the Fourier transform
of $f$ with a random phase. We have 
\begin{eqnarray*}
\langle h(\mathbf{r}+\Delta\mathbf{r})h(\mathbf{r})\rangle & = & \frac{1}{2}\left\langle \left(\int e^{i\mathbf{k}\cdot(\mathbf{r}+\Delta\mathbf{r})}e^{i\theta(\mathbf{k})}f(\mathbf{k})d\mathbf{k}+\text{c.c}\right)\left(\int e^{i\mathbf{k}'\cdot\mathbf{r}}e^{i\theta(\mathbf{k}')}f(\mathbf{k}')d\mathbf{k}'+\text{c.c}\right)\right\rangle \\
 & = & \frac{1}{2}\left(\int\int\langle e^{i(\theta(\mathbf{k})+\theta(\mathbf{k}'))}\rangle e^{i(\mathbf{k}\cdot(\mathbf{r}+\Delta\mathbf{r})+\mathbf{k}'\cdot\mathbf{r})}f(\mathbf{k})f(\mathbf{k}')d\mathbf{k}d\mathbf{k}'+\text{c.c}\right)\\
 &  & +\frac{1}{2}\left(\int\int\langle e^{i(\theta(\mathbf{k})-\theta(\mathbf{k}'))}\rangle e^{i(\mathbf{k}\cdot(\mathbf{r}+\Delta\mathbf{r})-\mathbf{k}'\cdot\mathbf{r})}f(\mathbf{k})\bar{f}(\mathbf{k}')d\mathbf{k}d\mathbf{k}'+\text{c.c}\right)\\
 & = & \int\cos(\mathbf{k}\cdot\Delta\mathbf{r})|f(\mathbf{k})|^{2}d\mathbf{k}\\
 & = & \text{Re}\left[\int f(\mathbf{r}+\Delta\mathbf{r})\bar{f}(\mathbf{r})d\mathbf{r}\right]
\end{eqnarray*}
which is just the real part of the convolution of $f$. In going to
line 4, we have made use of the fact that $\langle e^{i(\theta(\mathbf{k})+\theta(\mathbf{k}'))}\rangle=0$
and $\langle e^{i(\theta(\mathbf{k})-\theta(\mathbf{k}'))}\rangle=\delta_{\mathbf{k},\mathbf{k}'}$,
since $\theta$ is uncorrelated with $\mathbf{k}$. Indeed, the random
phase 'randomizes' $f(\mathbf{r})$, replacing the convolution with
the autocorrelation function. This result can also be extended to
higher correlations of even orders. 
\end{widetext}

If given a desired correlation function $\langle h(\mathbf{r})h(\mathbf{0})\rangle$,
one can find the requisite initial distribution via 
\begin{equation}
f(\mathbf{r})=\int e^{i\mathbf{k}\cdot\mathbf{r}}\sqrt{\int e^{-i\mathbf{k}\cdot\mathbf{r}'}\langle h(\mathbf{r})h(\mathbf{0})\rangle d\mathbf{r}'}d\mathbf{k}
\end{equation}
In this paper, the Gaussian distribution has the special property
that its autocorrelation is still a Gaussian, albeit with twice the
variance: 
\[
\int e^{-(x+\Delta x)^{2}}e^{-x^{2}}dx=e^{-\frac{(\Delta x)^{2}}{2}}
\]

\section{Calculation of mode transmission coefficient and local density of
states}

Our computation of the transmission coefficient is adapted from the
extension of the commonly used nonequilibrium Green's function method
described in Ong and Zhang~\cite{ZYOng:PRB15_Efficient}. Although
the approach was originally proposed for the study of nanoscale interfacial
phonon transmission, the structure of the equations describing the
topological lattice waveguide is in fact identical to that of the
equations typically used to model nanoscale phonons and is thus compatible
with the method, allowing us to apply the method to the macroscopic
topological phononic lattice system.

In the method, the one-dimensional (1D) system is divided into three
parts: the left lead, the central scattering region and the right
lead. The finite width of the leads means that the waveguide can be
treated as a multichannel system. At each frequency, a propagating
mode in the left lead is treated as a transmitting channel while a
propagating mode in the right lead is a receiving channel. Numerically,
the system identifies and extracts the eigenmodes of the left and
right lead from the uncoupled surface Green's function of the respective
leads. The retarded Green's function relating the left and right edges
of the scattering region is also computed and then used to calculate
the transition amplitude between each transmitting channel and each
receiving channel. Formally, this is equivalent to calculating the
scattering amplitude between the left lead modes and the right lead
modes.

\subsection{Structure and geometry of lattice waveguide}

\begin{figure*}
\includegraphics[width=14cm]{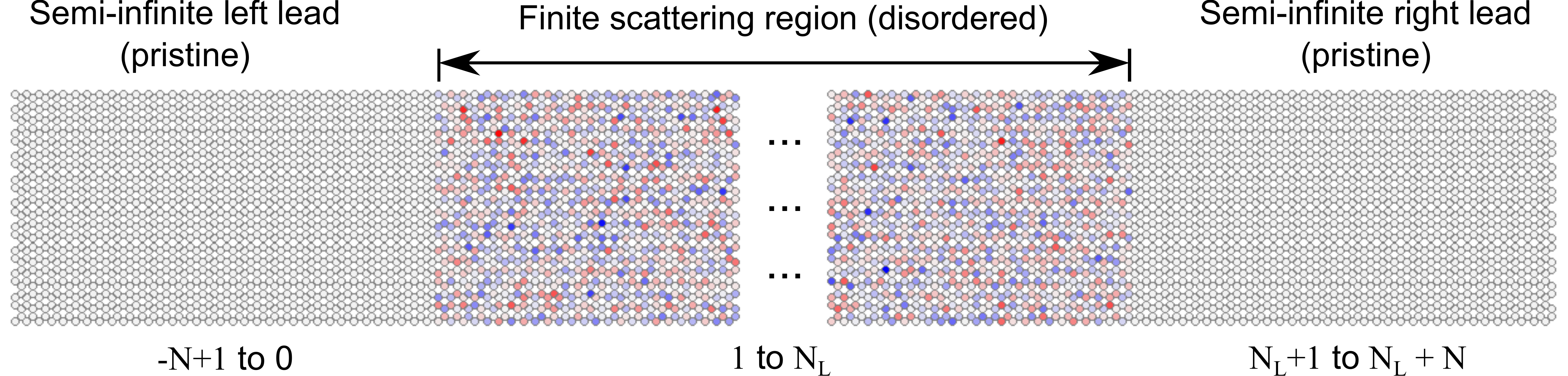}

\caption{Schematic of the different components of the lattice waveguide: the
left lead, the scattering region and the right lead. The enumeration
of the cells in each component is also given.}
\label{Fig:NEGF_Setup} 
\end{figure*}

The width of the waveguide is $N_{W}=20$ unit cells across and its
length is $2N+N_{L}$ unit cells. Hence, each cell (or principal layer)
of the waveguide consists of $20$ unit cells. As shown in Fig.~\ref{Fig:NEGF_Setup},
the waveguide can be divided into three components: the pristine left
lead, the scattering region with disorder and the pristine right lead.
There are $N$ cells in each lead and $N_{L}$ cells in the scattering
region. We enumerate the cells from $n=N^{-}$ to $n=N^{+}$ where
$N^{-}=-N+1$ and $N^{+}=N_{L}+N$ . Although we will take the limit
$N\rightarrow\infty$ eventually, we treat $N$ as a finite number
in the following description of the setup of the matrices and equations.

\subsection{Equation of motion}
\begin{widetext}
The equation of motion for the entire lattice waveguide can be compactly
written as 
\begin{equation}
\left(\mathbf{M}\frac{d^{2}}{dt^{2}}-\mathbf{K}\right)\mathbf{U}=\mathbf{0}\label{Eq:EqnOfMotion}
\end{equation}
where $\mathbf{K}$ and $\mathbf{M}$ are the stiffness and mass matrices,
respectively, and $\mathbf{U}$ is the column vector of displacement
coordinates. The stiffness matrix in Eq.~(\ref{Eq:EqnOfMotion})
can be written in the block-tridiagonal form with diagonal and off-diagonal
bands of submatrices,  
\begin{equation}
\mathbf{K}=\left(\begin{array}{cccccc}
\boldsymbol{K}_{N^{-},N^{-}} & \boldsymbol{K}_{N^{-},N^{-}+1}\\
\boldsymbol{K}_{N^{-}+1,N^{-}} & \ddots & \ddots\\
 & \ddots & \boldsymbol{K}_{n,n} & \boldsymbol{K}_{n,n+1}\\
 &  & \boldsymbol{K}_{n+1,n} & \ddots & \ddots\\
 &  &  & \ddots & \boldsymbol{K}_{N^{+}-1,N^{+}-1} & \boldsymbol{K}_{N^{+}-1,N^{+}}\\
 &  &  &  & \boldsymbol{K}_{N^{+},N^{+}-1} & \boldsymbol{K}_{N^{+},N^{+}}
\end{array}\right)\label{Eq:BigStiffnessMatrix}
\end{equation}
where each $\boldsymbol{K}_{n,m}$ is an $80\times80$ submatrix and
$n$ ($m=n\pm1$) is the column (row) \emph{group} index representing
the position of the cell. Since the spring coupling between adjacent
cells is identical, we set $\boldsymbol{K}_{n,n}=\boldsymbol{K}_{0,0}$,
$\boldsymbol{K}_{n,n+1}=\boldsymbol{K}_{0,1}$ and $\boldsymbol{K}_{n,n-1}=\boldsymbol{K}_{1,0}$
for $n=N^{-},\ldots,N^{+}$. In addition, it follows from the Hermiticity
of $\mathbf{K}$ that $\boldsymbol{K}_{n+1,n}=(\boldsymbol{K}_{n,n+1})^{\dagger}$.

\end{widetext}

The mass matrix in Eq.~(\ref{Eq:EqnOfMotion}) can be expressed as
\begin{equation}
\mathbf{M}=\left(\begin{array}{ccccc}
\boldsymbol{M}_{N^{-}}\\
 & \ddots\\
 &  & \boldsymbol{M}_{1}\\
 &  &  & \ddots\\
 &  &  &  & \boldsymbol{M}_{N^{+}}
\end{array}\right)\label{Eq:BigMassMatrix}
\end{equation}
where $\boldsymbol{M}_{n}$ is an $80\times80$ submatrix representing
the effective mass of the $n$-th cell for $n=N^{-},\ldots,N^{+}$.
However, unlike Eq.~(\ref{Eq:BigStiffnessMatrix}), the masses associated
with each cell are not necessarily periodic although the submatrices
for the pristine left lead ($N^{-}\leq n\leq0$) and right lead ($N_{L}+1\leq n\leq N^{+}$)
are identical,\emph{i.e.}, $\boldsymbol{M}_{n}=\boldsymbol{M}_{0}$
for $N^{-}\leq n<1$ and $N_{L}<n\leq N^{+}$. The submatrices in
the central scattering region ($1\leq n\leq N_{L}$) are however not
identical because of mass disorder. Each of the submatrices $\boldsymbol{M}_{n}$
in the central scattering region ($1\leq n\leq N_{L}$) can be written
in the block-diagonal form 
\begin{equation}
\boldsymbol{M}_{n}=\left(\begin{array}{ccc}
\boldsymbol{m}(\mathbf{r}_{1})\\
 & \ddots\\
 &  & \boldsymbol{m}(\mathbf{r}_{40})
\end{array}\right)\label{Eq:BigMassSubmatrix}
\end{equation}
where $\mathbf{r}_{l}$ is the position of $l$-th mass within the
cell, and 
\begin{equation}
\boldsymbol{m}(\mathbf{r})=[\langle m\rangle+\delta m(\mathbf{r})]\left(\begin{array}{cc}
1 & -i\alpha\\
i\alpha & 1
\end{array}\right)
\end{equation}
is the $2\times2$ matrix representing the effective mass at $\mathbf{r}$
with $\alpha=0.3$ like in Ref. \cite{PWang:PRL15_Topological}.
\begin{widetext}
Equation~(\ref{Eq:EqnOfMotion}) can be simplified to the matrix
equation that is second order in time, \emph{i.e.} 
\begin{equation}
\left(\frac{d^{2}}{dt^{2}}+\mathbf{H}\right)\mathbf{V}=\mathbf{0}\label{Eq:AlternateEqnOfMotion}
\end{equation}
where $\mathbf{V=\mathbf{M}}^{1/2}\mathbf{U}$, and $\mathbf{H}=\mathbf{M}^{-1/2}\mathbf{K}\mathbf{M}^{-1/2}$
is the mass-normalized force constant matrix that can be written in
the block-tridiagonal form 
\begin{equation}
\mathbf{H}=\left(\begin{array}{cccccc}
\boldsymbol{H}_{N^{-},N^{-}} & \boldsymbol{H}_{N^{-},N^{-}+1}\\
\boldsymbol{H}_{N^{-}+1,N^{-}} & \ddots & \ddots\\
 & \ddots & \boldsymbol{H}_{1,1} & \boldsymbol{H}_{1,2}\\
 &  & \boldsymbol{H}_{2,1} & \ddots & \ddots\\
 &  &  & \ddots & \boldsymbol{H}_{N^{+}-1,N^{+}-1} & \boldsymbol{H}_{N^{+}-1,N^{+}}\\
 &  &  &  & \boldsymbol{H}_{N^{+},N^{+}-1} & \boldsymbol{H}_{N^{+},N^{+}}
\end{array}\right)\ .\label{Eq:Hamiltonian}
\end{equation}
Each submatrix $\boldsymbol{H}_{n,m}$ is given by $\boldsymbol{H}_{n,m}=\boldsymbol{M}_{n}^{-1/2}\boldsymbol{K}_{n,m}\boldsymbol{M}_{m}^{-1/2}$
where $m=n$ or $n\pm1$. The Hermiticity of $\mathbf{H}$ implies
that $\boldsymbol{H}_{n,m}=(\boldsymbol{H}_{m,n})^{\dagger}$.
\end{widetext}

\subsection{Division of waveguide into scattering region and leads}
\begin{widetext}
We recall in Fig.~\ref{Fig:NEGF_Setup} that the scattering region
corresponds to the cells for $1\leq n\leq N_{L}$ while the left (right)
lead corresponds to the cells for $N^{-}\leq n\leq0$ ($N_{L}+1\leq n\leq N^{+}$).
In the left and right lead, there is no mass disorder or $\delta m(\mathbf{r})=0$
while in the scattering region, it is determined by Eq.~(\ref{Eq:RandomMassDisorder}).
We can write Eq.~(\ref{Eq:Hamiltonian}) as  
\begin{equation}
\mathbf{H}=\left(\begin{array}{ccc}
\mathbf{H}_{L} & \mathbf{H}_{LC}\\
\mathbf{H}_{CL} & \mathbf{H}_{C} & \mathbf{H}_{CR}\\
 & \mathbf{H}_{RC} & \mathbf{H}_{R}
\end{array}\right)\label{Eq:TotalHamiltonian}
\end{equation}
where \begin{subequations} 
\begin{equation}
\mathbf{H}_{L}=\left(\begin{array}{cccc}
\boldsymbol{H}_{N_{-},N_{-}} & \boldsymbol{H}_{N_{-},N_{-}+1}\\
\boldsymbol{H}_{N_{-}+1,N_{-}} & \ddots & \ddots\\
 & \ddots & \boldsymbol{H}_{-1,-1} & \boldsymbol{H}_{-1,0}\\
 &  & \boldsymbol{H}_{0,-1} & \boldsymbol{H}_{0,0}
\end{array}\right)\label{Subeq:HL_matrix}
\end{equation}

\begin{equation}
\mathbf{H}_{C}=\left(\begin{array}{cccc}
\boldsymbol{H}_{1,1} & \boldsymbol{H}_{1,2}\\
\boldsymbol{H}_{2,1} & \boldsymbol{H}_{2,2} & \ddots\\
 & \ddots & \ddots & \boldsymbol{H}_{N_{L}-1,N_{L}-1}\\
 &  & \boldsymbol{H}_{N_{L},N_{L}-1} & \boldsymbol{H}_{N_{L},N_{L}}
\end{array}\right)\label{Subeq:HC_matrix}
\end{equation}
\begin{equation}
\mathbf{H}_{R}=\left(\begin{array}{cccc}
\boldsymbol{H}_{N_{L}+1,N_{L}+1} & \boldsymbol{H}_{N_{L}+1,N_{L}+2}\\
\boldsymbol{H}_{N_{L}+2,N_{L}+1} & \boldsymbol{H}_{N_{L}+2,N_{L}+2} & \ddots\\
 & \ddots & \ddots & \boldsymbol{H}_{N_{+}-1,N_{+}}\\
 &  & \boldsymbol{H}_{N_{+},N_{+}-1} & \boldsymbol{H}_{N_{+},N_{+}}
\end{array}\right)\label{Subeq:HR_matrix}
\end{equation}
\end{subequations} The matrices in Eqs.~(\ref{Subeq:HL_matrix}),
(\ref{Subeq:HC_matrix}) and (\ref{Subeq:HR_matrix}) correspond to
the left lead, the scattering region and the right lead, respectively.
The periodicity in the arrangement of the stiffness and mass matrices
of the left lead implies that the diagonal and off-diagonal submatrices
of Eq.~(\ref{Subeq:HL_matrix}) satisfy the following conditions
\[
\boldsymbol{H}_{N_{-},N_{-}}=\boldsymbol{H}_{N_{-}+1,N_{-}+1}=\ldots=\boldsymbol{H}_{0,0}
\]
\[
\boldsymbol{H}_{N_{-},N_{-}+1}=\boldsymbol{H}_{N_{-}+1,N_{-}+2}=\ldots=\boldsymbol{H}_{-1,0}
\]
\[
\boldsymbol{H}_{N_{-}+1,N_{-}}=\boldsymbol{H}_{N_{-}+2,N_{-}+1}=\ldots=\boldsymbol{H}_{0,-1}
\]

\end{widetext}

Likewise, the diagonal and off-diagonal submatrices of Eq.~(\ref{Subeq:HR_matrix})
also satisfy 
\[
\boldsymbol{H}_{N_{L}+1,N_{L}+1}=\ldots=\boldsymbol{H}_{N_{+},N_{+}}=\boldsymbol{H}_{0,0}
\]
\[
\boldsymbol{H}_{N_{L}+1,N_{L}+2}=\ldots=\boldsymbol{H}_{N_{+}-1,N_{+}}=\boldsymbol{H}_{-1,0}
\]
\[
\boldsymbol{H}_{N_{L}+2,N_{L}+1}=\ldots=\boldsymbol{H}_{N_{+},N_{+}-1}=\boldsymbol{H}_{0,-1}
\]
Given the uniformity in the stiffness and mass submatrices, the dispersion
($\omega-\mu$) relationship for the propagating modes of the semi-infinite
leads is determined by solving the equation 
\begin{equation}
\det(\boldsymbol{H}_{0,-1}\lambda^{-1}+\boldsymbol{H}_{0,0}+\boldsymbol{H}_{0-1,0}\lambda-\omega^{2}\boldsymbol{I})=0\label{Eq:PhononDispersion}
\end{equation}
where $\boldsymbol{I}$ is the identity matrix and $\lambda=e^{i\mu a}$
is the Bloch factor. $\omega$ and $\mu$ are respectively the frequency
and the wave vector. In the scattering region ($1\leq n\leq N_{L}$),
the mass at each lattice site $m(\mathbf{r})=\langle m\rangle+\delta m(\mathbf{r})$
varies with the position $\mathbf{r}$ according to Eq.~(\ref{Eq:RandomMassDisorder}).
The mass disorder of the scattering region causes the left lead propagating
modes to be partially transmitted to the right lead.

\subsection{Green's functions for leads and scattering region }

In the frequency domain, Eq.~(\ref{Eq:TotalHamiltonian}) becomes
\begin{equation}
(\omega^{2}\mathbf{I}-\mathbf{H})\tilde{\mathbf{V}}=\mathbf{0}\label{Eq:FrequencyEqnOfMotion}
\end{equation}
where $\tilde{\mathbf{V}}$ is the Fourier transform of $\mathbf{V}$
and $\mathbf{I}$ is the identity matrix. The retarded Green's function
corresponding to the linear operator in Eq.~(\ref{Eq:FrequencyEqnOfMotion})
is 
\begin{equation}
\mathbf{G}^{\text{ret}}(\omega)=[(\omega^{2}+i0^{+})\mathbf{I}-\mathbf{H}]^{-1}\label{Eq:TotalRetardedGreensFn}
\end{equation}
and Eq.~(\ref{Eq:TotalRetardedGreensFn}) can be expressed as

\[
\mathbf{G}^{\text{ret}}(\omega)=\left(\begin{array}{ccc}
\mathbf{G}_{L}^{\text{ret}}(\omega) & \mathbf{G}_{LC}^{\text{ret}}(\omega) & \mathbf{G}_{LR}^{\text{ret}}(\omega)\\
\mathbf{G}_{CL}^{\text{ret}}(\omega) & \mathbf{G}_{C}^{\text{ret}}(\omega) & \mathbf{G}_{CR}^{\text{ret}}(\omega)\\
\mathbf{G}_{RL}^{\text{ret}}(\omega) & \mathbf{G}_{RC}^{\text{ret}}(\omega) & \mathbf{G}_{R}^{\text{ret}}(\omega)
\end{array}\right)\ .
\]
In order to compute the left lead mode transmission coefficients,
we first need to find the retarded Green's function $\mathbf{G}_{C}^{\text{ret}}(\omega)$
for the finite scattering region. It is given by the expression 
\begin{equation}
\mathbf{G}_{C}^{\text{ret}}(\omega)=[(\omega^{2}+i0^{+})\mathbf{I}_{C}-\mathbf{H}_{C}^{\textrm{eff}}(\omega)]^{-1}\ ,\label{Eq:CenterGreensFn}
\end{equation}
where the term $\mathbf{H}^{\textrm{eff}}(\omega)$ is the $\omega$-dependent
effective Hamiltonian 
\begin{equation}
\mathbf{H}_{C}^{\textrm{eff}}(\omega)=\mathbf{H}_{C}+\mathbf{H}_{CL}\mathbf{g}_{L}^{\text{ret}}(\omega)\mathbf{H}_{LC}+\mathbf{H}_{CR}\mathbf{g}_{R}^{\text{ret}}(\omega)\mathbf{H}_{RC}\label{Eq:EffectiveHamiltonian}
\end{equation}
with 
\[
\mathbf{g}_{L}^{\text{ret}}(\omega)=[(\omega^{2}+i0^{+})\mathbf{I}_{L}-\mathbf{H}_{L}]^{-1}
\]
\[
\mathbf{g}_{R}^{\text{ret}}(\omega)=[(\omega^{2}+i0^{+})\mathbf{I}_{R}-\mathbf{H}_{R}]^{-1}
\]
The LHS of Eq.~(\ref{Eq:EffectiveHamiltonian}) can be written more
explicitly as 
\begin{equation}
\mathbf{H}_{C}^{\textrm{eff}}(\omega)=\left(\begin{array}{cccc}
\boldsymbol{H}_{1,1}^{\textrm{eff}} & \boldsymbol{H}_{1,2}\\
\boldsymbol{H}_{2,1} & \boldsymbol{H}_{2,2} & \ddots\\
 & \ddots & \ddots & \boldsymbol{H}_{N_{L}-1,N_{L}}\\
 &  & \boldsymbol{H}_{N_{L},N_{L}-1} & \boldsymbol{H}_{N_{L},N_{L}}^{\textrm{eff}}
\end{array}\right)\ ,\label{Eq:EffectiveHamiltonianMatrix}
\end{equation}
with 
\[
\boldsymbol{H}_{1,1}^{\textrm{eff}}=\boldsymbol{H}_{1,1}+\boldsymbol{H}_{1,0}\boldsymbol{g}_{0,0}^{\text{ret}}(\omega)\boldsymbol{H}_{0,1}
\]
and 
\[
\boldsymbol{H}_{N_{L},N_{L}}^{\textrm{eff}}=\boldsymbol{H}_{N_{L},N_{L}}+\boldsymbol{H}_{N_{L},N_{L}+1}\boldsymbol{g}_{N_{L}+1,N_{L}+1}^{\text{ret}}(\omega)\boldsymbol{H}_{N_{L}+1,N_{L}}\ .
\]
The matrices $\boldsymbol{g}_{0,0}^{\text{ret}}(\omega)$ and $\boldsymbol{g}_{N_{L}+1,N_{L}+1}^{\text{ret}}(\omega)$
represent the surface Green's function of the uncoupled left and right
lead, respectively, and in the limit $N\rightarrow\infty$ where the
leads become infinitely large, they satisfy the equations \begin{subequations}
\begin{align}
\boldsymbol{g}_{0,0}^{\text{ret}}(\omega)= & [(\omega^{2}+i0^{+})\boldsymbol{I}-\boldsymbol{H}_{0,0}\nonumber \\
 & -\boldsymbol{H}_{0,-1}\boldsymbol{g}_{0,0}^{\text{ret}}(\omega)\boldsymbol{H}_{-1,0}]^{-1}\label{Subeq:LeftSurfGreensFn}
\end{align}
\begin{align}
\boldsymbol{g}_{N_{L}+1,N_{L}+1}^{\text{ret}}(\omega)= & [(\omega^{2}+i0^{+})\boldsymbol{I}-\boldsymbol{H}_{0,0}\nonumber \\
 & -\boldsymbol{H}_{-1,0}\boldsymbol{g}_{N_{L}+1,N_{L}+1}^{\text{ret}}(\omega)\boldsymbol{H}_{0,-1}]^{-1}\ .\label{Subeq:RightSurfGreensFn}
\end{align}
\label{Eq:AllSurfGreensFns}\end{subequations} The nonlinear equations
in Eq.~(\ref{Eq:AllSurfGreensFns}) can be solved numerically with
the decimation technique~\cite{ZYOng:PRB15_Efficient} to yield the
surface Green's functions $\boldsymbol{g}_{0,0}^{\text{ret}}(\omega)$
and $\boldsymbol{g}_{N_{L}+1,N_{L}+1}^{\text{ret}}(\omega)$.

\subsection{Surface Green's functions, Bloch matrices and eigenmodes}

It is shown by Ong and Zhang~\cite{ZYOng:PRB15_Efficient} that the
constant-$\omega$ propagating and evanescent eigenmodes can be extracted
from the surface Green's functions. We first compute the corresponding
Bloch matrices \begin{subequations} 
\begin{equation}
\boldsymbol{F}_{L}^{\text{adv}}(-)^{-1}=[\boldsymbol{H}_{0,-1}\boldsymbol{g}_{0,0}^{\text{ret}}]^{\dagger}\label{Subeq:LeftBlochMatrix}
\end{equation}
and 
\begin{equation}
\boldsymbol{F}_{R}^{\text{ret}}(+)=\boldsymbol{g}_{N_{L}+1,N_{L}+1}^{\text{ret}}\boldsymbol{H}_{0,-1}\ .\label{Subeq:RightBlochMatrix}
\end{equation}
\label{Eq:BlochMatrices}\end{subequations} The matrices $\boldsymbol{U}_{L}^{\text{adv}}(-)$
and $\boldsymbol{U}_{R}^{\text{ret}}(+)$, in which the column vectors
represent the extended and evanescent eigenmodes, are obtained by
solving numerically the equations \begin{subequations} 
\begin{equation}
\boldsymbol{F}_{L}^{\text{adv}}(-)\boldsymbol{U}_{L}^{\text{adv}}(-)=\boldsymbol{U}_{L}^{\text{adv}}(-)\boldsymbol{\Lambda}_{L}^{\text{adv}}(-)\label{Subeq:LeftBlochEigen}
\end{equation}
and 
\begin{equation}
\boldsymbol{F}_{R}^{\text{ret}}(+)\boldsymbol{U}_{R}^{\text{ret}}(+)=\boldsymbol{U}_{R}^{\text{ret}}(+)\boldsymbol{\Lambda}_{R}^{\text{ret}}(+)\ .\label{Subeq:RightBlochEigen}
\end{equation}
\label{Eq:BlochEigeneqns}\end{subequations} The matrices $\boldsymbol{\Lambda}_{L}^{\text{adv}}(-)$
and $\boldsymbol{\Lambda}_{R}^{\text{ret}}(+)$ are diagonal matrices
with the diagonal elements equal to the Bloch factor $\lambda=\exp(\mp i\mu a)$
of the corresponding propagating eigenmodes, where $a$ is the one-dimensional
lattice constant. 
\[
\boldsymbol{\Lambda}_{L}^{\text{adv}}(-)=\left(\begin{array}{cccc}
\lambda_{1} & 0\\
0 & \lambda_{2} & \ddots\\
 & \ddots & \ddots & 0\\
 &  & 0 & \lambda_{80}
\end{array}\right)
\]
Thus, the wave vector $\mu$ of the eigenmode can be easily determined
from $\lambda$.

The group velocity matrices for the eigenmodes are \begin{subequations}
\begin{align}
\boldsymbol{V}_{L}(+)= & \frac{ia}{2\omega}\boldsymbol{U}_{L}^{\text{adv}}(-)^{\dagger}\boldsymbol{H}_{0,-1}\nonumber \\
 & \times[\boldsymbol{g}_{0,0}^{\text{ret}}(\omega)-\boldsymbol{g}_{0,0}^{\text{ret}}(\omega)^{\dagger}]\boldsymbol{H}_{-1,0}\boldsymbol{U}_{L}^{\text{adv}}(-)\label{Subeq:LeftVelocityMatrix}
\end{align}
and 
\begin{align}
\boldsymbol{V}_{R}(+)= & \frac{ia}{2\omega}\mathbf{U}_{R}^{\text{ret}}(+)^{\dagger}\boldsymbol{H}_{-1,0}[\boldsymbol{g}_{N+1,N+1}^{\text{ret}}(\omega)\nonumber \\
 & -\boldsymbol{g}_{N+1,N+1}^{\text{ret}}(\omega)^{\dagger}]\boldsymbol{H}_{0,-1}\boldsymbol{U}_{R}^{\text{ret}}(+)\ ,\label{Subeq:RightVelocityMatrix}
\end{align}
\label{Eq:VelocityMatrices}\end{subequations} which are needed for
calculating the transmission coefficient later. The off-diagonal elements
of the velocity matrices in Eq.~\ref{Eq:VelocityMatrices} are zero
while the diagonal elements are positive only if the corresponding
eigenmode is propagating (extended).

Like $\mathbf{H}_{C}^{\textrm{eff}}(\omega)$ in Eq.~(\ref{Eq:EffectiveHamiltonianMatrix}),
the matrix $\mathbf{G}_{C}^{\text{ret}}(\omega)$ in Eq.~(\ref{Eq:CenterGreensFn})
can be written in the block form 
\[
\mathbf{G}_{C}^{\text{ret}}(\omega)=\left(\begin{array}{cccc}
\boldsymbol{G}_{1,1}^{\textrm{ret}} & \boldsymbol{G}_{1,2}^{\textrm{ret}} & \cdots & \boldsymbol{G}_{1,N_{L}}^{\textrm{ret}}\\
\boldsymbol{G}_{2,1}^{\textrm{ret}} & \boldsymbol{G}_{2,2}^{\textrm{ret}} & \ddots & \vdots\\
\vdots & \ddots & \ddots & \boldsymbol{G}_{N_{L}-1,N_{L}}^{\textrm{ret}}\\
\boldsymbol{G}_{N_{L},1}^{\textrm{ret}} & \cdots & \boldsymbol{G}_{N_{L},N_{L}-1}^{\textrm{ret}} & \boldsymbol{G}_{N_{L},N_{L}}^{\textrm{ret}}
\end{array}\right)\ .
\]
The submatrix $\boldsymbol{G}_{n,m}^{\textrm{ret}}(\omega)$ can be
interpreted as the frequency-domain transfer function for the vibrational
response at unit cell $n$ to an oscillatory harmonic driving force
at unit cell $m$ with frequency $\omega$. Intuitively, information
on energy transfer from the leftmost cell to the rightmost cell of
the scattering region should be contained in the submatrix $\boldsymbol{G}_{N_{L},1}^{\textrm{ret}}(\omega)$.

\subsection{Transmission amplitude matrix}

The transmission amplitude matrix is 
\begin{align}
\boldsymbol{t}(\omega)= & \frac{2i\omega}{a}\boldsymbol{V}_{R}(+)^{\nicefrac{1}{2}}[\boldsymbol{U}_{R}^{\text{ret}}(+)]^{-1}\nonumber \\
 & \times\boldsymbol{\mathcal{G}}_{C}^{\text{ret}}(\omega)[\boldsymbol{U}_{L}^{\text{adv}}(-)^{\dagger}]^{-1}\boldsymbol{V}_{L}(+)^{\nicefrac{1}{2}}\label{Eq:TransmissionMatrix}
\end{align}
where 
\[
\boldsymbol{\mathcal{G}}_{C}^{\text{ret}}(\omega)=\boldsymbol{g}_{N_{L}+1,N_{L}+1}^{\text{ret}}(\omega)\boldsymbol{H}_{0,-1}\boldsymbol{G}_{N_{L},1}^{\textrm{ret}}(\omega)\boldsymbol{H}_{0,-1}\boldsymbol{g}_{0,0}^{\text{ret}}(\omega)\ .
\]
 The individual matrix elements of the transmission matrix in Eq.~(\ref{Eq:TransmissionMatrix})
gives the transition probability amplitude between an in-coming left
lead channel and an out-going right lead channel at frequency $\omega$.
For instance, $t_{mn}$ gives the transition probability amplitude
between the in-coming left lead eigenmode corresponding to the $n$-th
column vector of $\boldsymbol{U}_{L}^{\text{adv}}(-)$, with its Bloch
factor given by the $n$-th diagonal element of $\boldsymbol{\Lambda}_{L}^{\text{adv}}(-)$,
and the out-going right lead eigenmode corresponding to the $m$-th
column vector of $\boldsymbol{U}_{R}^{\text{ret}}(+)$ with its Bloch
factor given by the $m$-th diagonal element of $\boldsymbol{\Lambda}_{R}^{\text{ret}}(+)$.
If either one of the eigenmodes is an evanescent mode, then its group
velocity is 0, \emph{i.e.} $[\boldsymbol{V}_{L}(+)]_{n,n}=0$ or $[\boldsymbol{V}_{R}(+)]_{m,m}=0$,
and the transition probability amplitude is $t_{mn}=0$. 

The transmission coefficient of the $n$-th left lead eigenmode $\Xi_{n}(\omega)$
is given by $\Xi_{n}(\omega)=\sum_{m}|t_{mn}(\omega)|^{2}$, or the
$n$-th diagonal element of $\boldsymbol{t}(\omega)^{\dagger}\boldsymbol{t}(\omega)$,
and has a numerical value between 0 and 1. The associated wavevector
$\mu$ can be determined by from the $n$-th diagonal element of $\boldsymbol{\Lambda}_{L}^{\text{adv}}(-)$
which yields the Bloch factor $\lambda=e^{i\mu a}$. The transmittance
at frequency $\omega$ can be computed from the sum of the transmission
coefficients, \emph{i.e.}, 
\begin{equation}
T(\omega)=\sum_{n}\Xi_{n}(\omega)=\text{Tr}[\boldsymbol{t}(\omega)^{\dagger}\boldsymbol{t}(\omega)]\ .\label{Eq:Transmittance}
\end{equation}
In the absence of any mass disorder in the scattering region, the
transmittance in Eq.~(\ref{Eq:Transmittance}) is an integer equal
to the number of channels in each lead at frequency $\omega$ since
$\Xi_{n}(\omega)=1$ for each propagating eigenmode and $0$ otherwise.

\subsection{Local density of states \label{SubAppend:LDOS}}

The local density of states (LDOS) at site $\mathbf{r}$ in the scattering
region is 
\begin{equation}
\rho(\omega,\mathbf{r})=\frac{i\omega}{\pi a}\sum_{n(\mathbf{r})}[\mathbf{G}_{C}^{\text{ret}}(\omega)-\mathbf{G}_{C}^{\text{ret}}(\omega)^{\dagger}]_{n(\mathbf{r}),n(\mathbf{r})}\label{Eq:LocalDOS}
\end{equation}
where $n(\mathbf{r})$ is the index of the degrees of freedom associated
with site $\mathbf{r}$. In effect, Eq. (\ref{Eq:LocalDOS}) is the
trace of a $2\times2$ matrix, and the LDOS has the units of inverse
length times inverse frequency. The scattering region has $N_{L}$
cells and each cell has $40$ lattice sites and $80$ degrees of freedom
since each lattice site has two degrees of freedom, one in $x$ and
the other in $y$. Hence, the entire scattering region has $80N_{L}$
degrees of freedom and $\mathbf{G}_{C}^{\text{ret}}$ is an $80N_{L}\times80N_{L}$
matrix. To find the LDOS at a site of a particular site $\mathbf{r}$,
we only need to sum over the two diagonal elements of $\mathbf{G}_{C}^{\text{ret}}$
corresponding to the $x$ and $y$ degrees of freedom at site $\mathbf{r}$.
In total, there are $40N_{L}$ lattice sites $\mathbf{r}$ and associated
LDOS values $\rho(\mathbf{r})$.

\section{Mode transmission coefficients for different correlation lengths
\label{Append:AdditionalTransmissionData}}

\begin{figure}
\includegraphics[width=8.5cm]{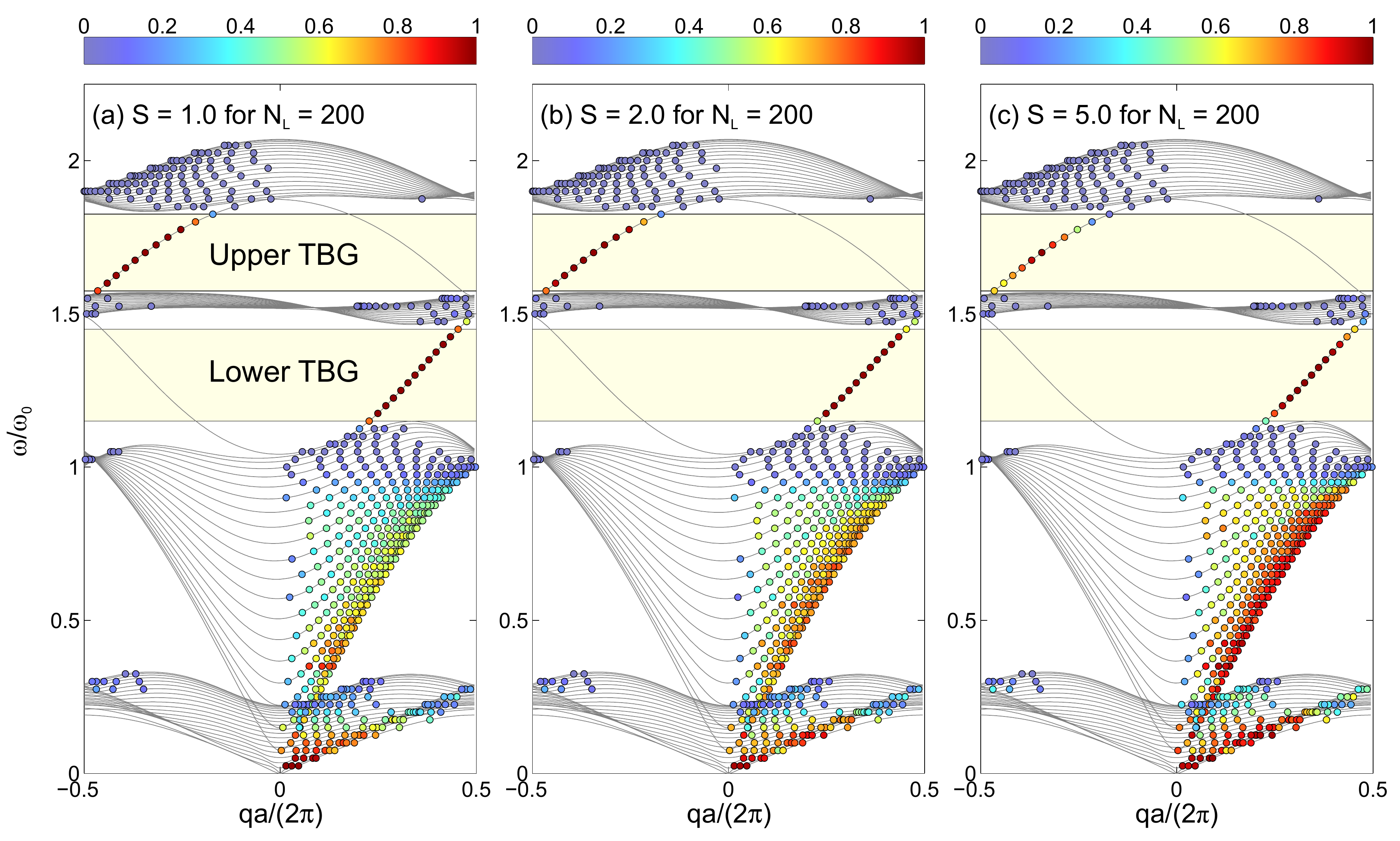}

\includegraphics[width=8.5cm]{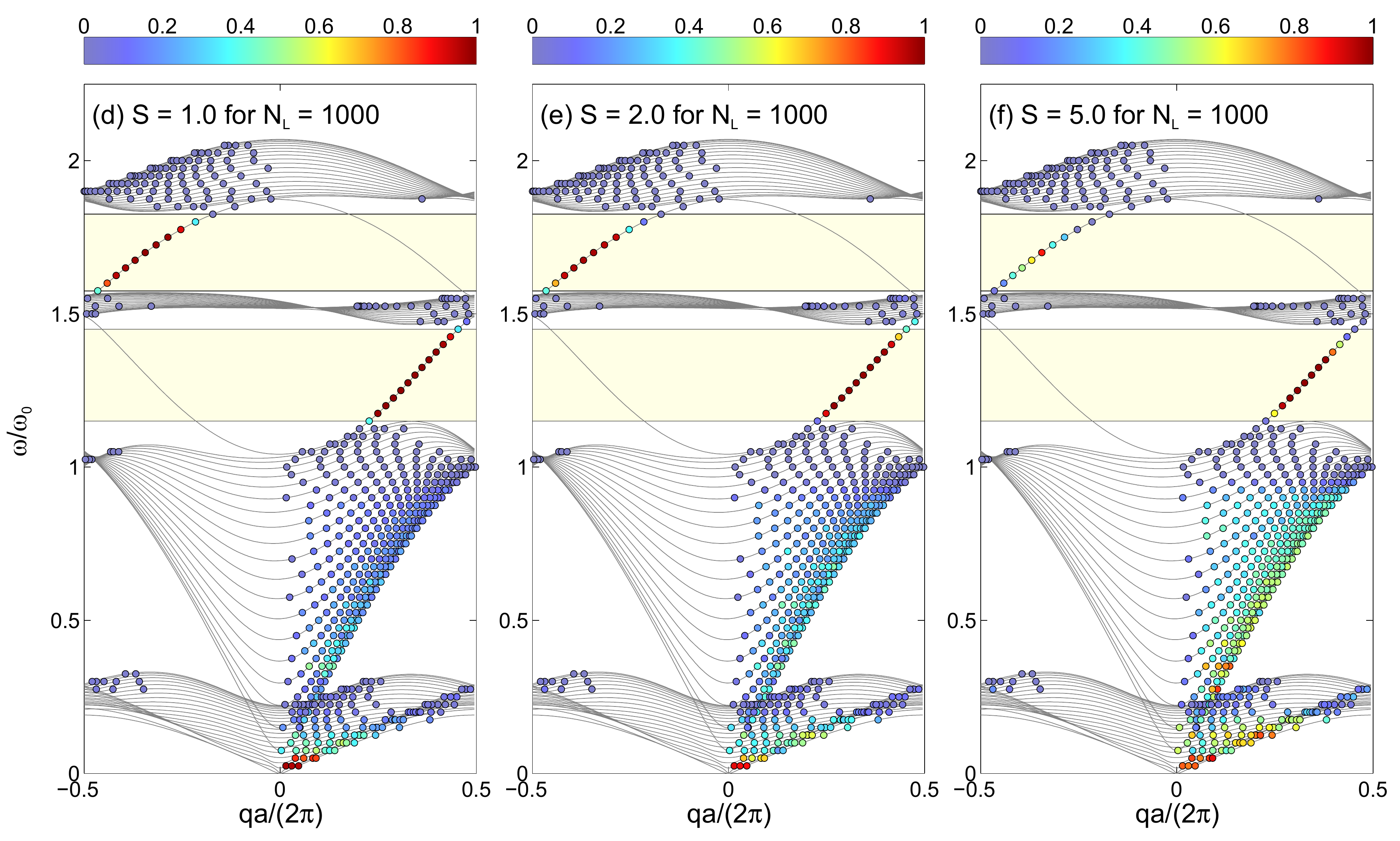}

\includegraphics[width=8.5cm]{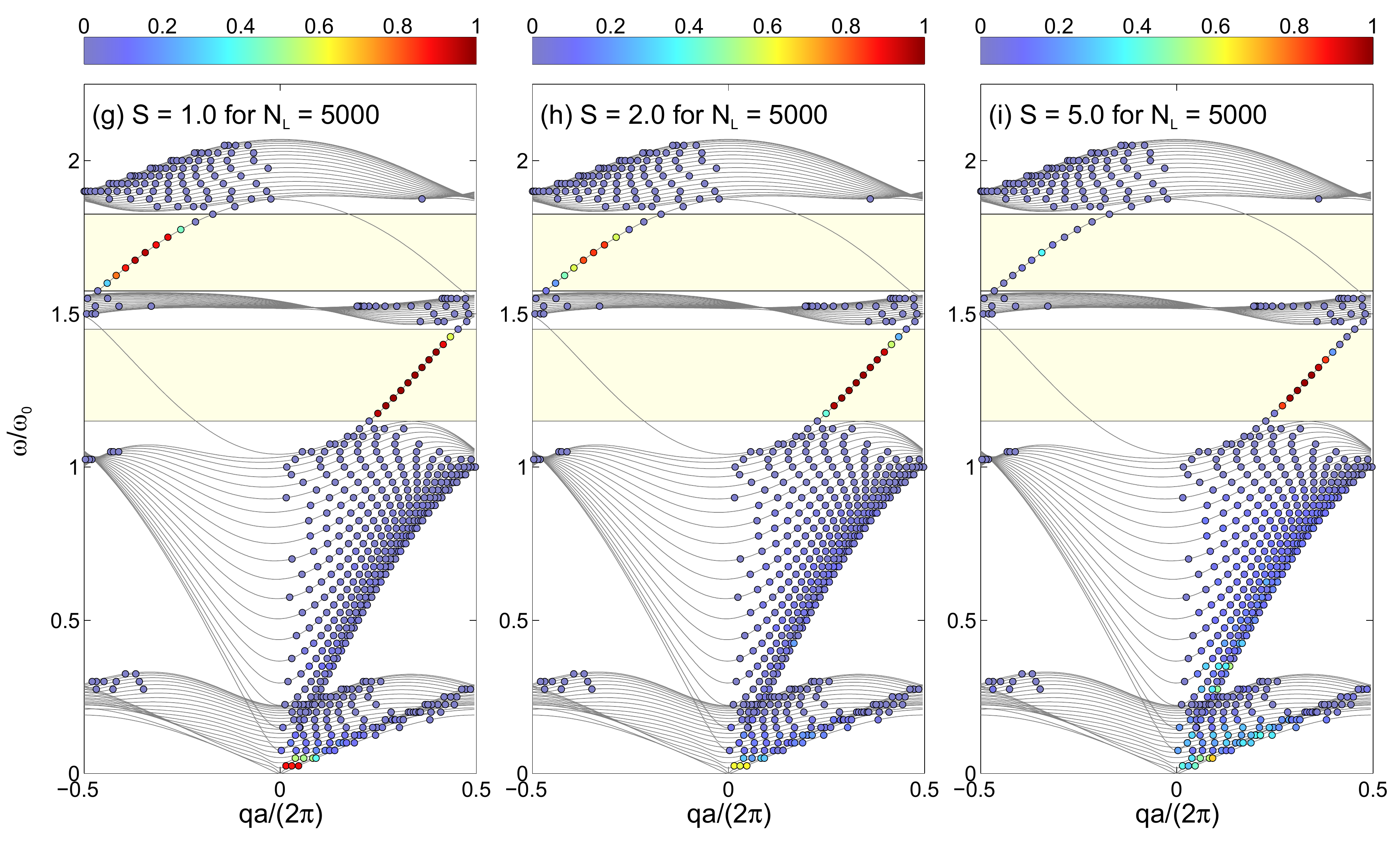}

\caption{Mode transmission coefficients (color in the small circles) for correlation
lengths $S=1.0$, $2.0$ and $5.0$ at (a-c) $N_{L}=200$, (d-f) $N_{L}=1000$
and (g-i) $N_{L}=5000$. The yellow-shaded regions correspond to the
topological band gaps. }
\label{Fig:ModeTransmission} 
\end{figure}

We plot the mode transmission coefficients in Fig.~\ref{Fig:ModeTransmission}
for different values of the correlation length $S$ at $N_{L}=200$,
$1000$ and $5000$. Figures~\ref{Fig:ModeTransmission}(g) to (i)
show that the mode transmission coefficients for the very low-frequency
modes and the TPE modes in the lower and upper topological band gap
(TBG) decrease as $S$ increases. On the other hand, the transmission
of the higher-frequency modes is enhanced as we increase $S$. 

\bibliographystyle{apsrev4-1}
\bibliography{references}

\end{document}